\documentclass[11pt,a4paper]{article}
\usepackage{jheppub}
\usepackage{amsmath,amssymb,graphicx,epstopdf,enumerate,amsfonts,booktabs,color}
\usepackage{float,placeins}
\providecommand{\U}[1]{\protect\rule{.1in}{.1in}}
\affiliation[a]{Institute of Physics, National Chiao Tung University, Hsinchu, ROC}
\affiliation[b]{Department of Electrophysics, National Chiao Tung University, Hsinchu, ROC}
\emailAdd{sr755332@gmail.com}
\emailAdd{yiyang@mail.nctu.edu.tw}
\emailAdd{phy.pro.phy@gmail.com}
\abstract{We study confinement-deconfinement phase transition in a holographic soft-wall QCD model. By solving the Einstein-Maxwell-scalar system analytically, we obtain the phase structure of the black hole backgrounds. We then impose probe open strings in such background to investigate the confinement-deconfinement phase transition from different open string configurations under various temperatures and chemical potentials. Furthermore, we study the Wilson loop by calculating the minimal surface of the probing open string world-sheet and obtain the Cornell potential in confinement phase analytically.}
\begin{document}

\title{Approaching Confinement Structure for Light Quarks in a Holographic Soft Wall QCD Model}
\author{Meng-Wei Li${^a}$, Yi Yang${^b}$ and Pei-Hung Yuan${^a}$}
\maketitle

\setcounter{equation}{0}
\renewcommand{\theequation}{\arabic{section}.\arabic{equation}}

\section{Introduction}
\begin{figure}[t!]
\begin{center}
\includegraphics[
height=2.05in, width=2.3in]
{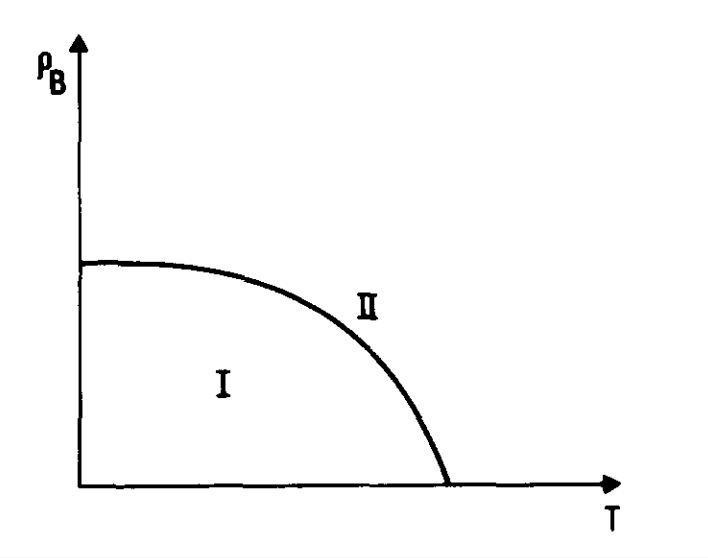}
\includegraphics[
height=2.1in, width=2.3in]
{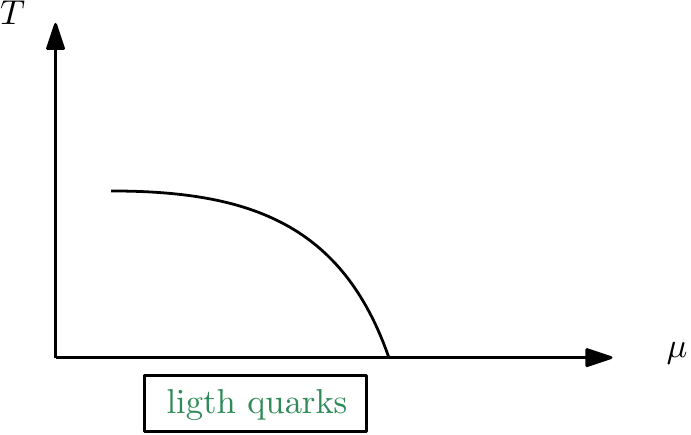}
\includegraphics[
height=2.1in, width=2.3in]
{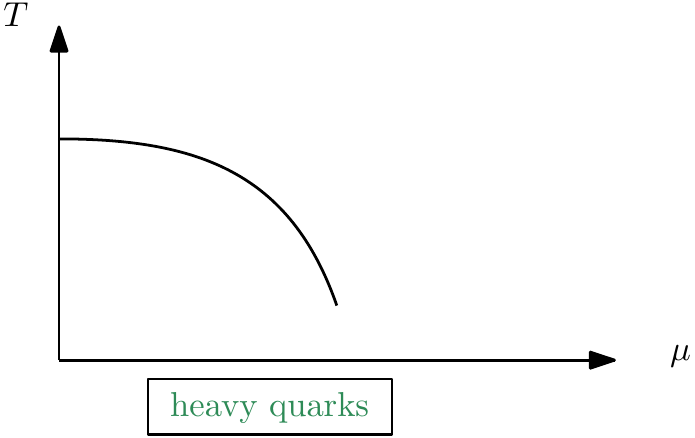}
\vskip -0.05cm \hskip 0.15 cm (a) \hskip 5.5 cm (b) \hskip 5.5 cm (c)
\end{center}
\caption{(a)The first schematic QCD phase diagram of hadronic matter claimed by Cabibbo and Parisi in 1975, where $\rho_B$ labels the baryon density. The transition line divided the phase space into two phases. Quarks are bounded in phase I and unconfined in phase II \cite{Cabibbo1975}. (b) Nowadays, lattice simulations convince people that the QCD phase diagram behaves as (a) in chiral limit, but exists a crossover regime for massive cases. Light quarks phase transition should display as (b), otherwise would present as (c) when considering heavy quarks even the pure gauge limit.} \label{fig_cabibbo1975}
\end{figure}

There are various phenomena of phase structure in Quantum Chromodynamics (QCD) theory. Confinement-deconfinement phase transition is one of the most important phenomena in QCD phase diagram. It is widely believed that, the system is in the confinement phase at low temperature $T$ and small chemical potential $\mu$ (low quark density) region, in which quarks are confined to hadronic bound states, e.g. mesons and baryons. While it is in the deconfinement phase at high temperature and large chemical potential (high quark density) region, in which free quarks exist, e.g. quark gluon plasma (QGP). It is natural to conjecture that there exists a phase transition between the two phases as showed in Fig.\ref{fig_cabibbo1975}(a) (carton phase diagram at m=0).

To understand the confinement-deconfinement phase transition in QCD is a very important but extremely difficult task. The interaction becomes strong around the phase transition region, so that the conventional perturbation method in quantum field theory does not work. For a long time, lattice QCD is the only method to attack the problem of phase transition in QCD \cite{1009.4089,1203.5320}. According to the calculation in lattice QCD, the confinement-deconfinement phase transition is first ordered in zero quark mass limit $m_q=0$. However, considering finite quark mass, part of the phase transition line would become crossover. For small and large quark masses, the behaviors of phase transformation at zero chemical potential $\mu=0$ have been calculated by lattice QCD. The phase diagrams are conjectured and showed in figure (carton phase diagrams at finite m) for light quark (b) and heavy quark (c), respectively.

However, lattice QCD faces the so-called sign problem at finite chemical potential. To study the full phase structure in QCD, we need new methods. During the last fifteen years, AdS/CFT duality has been developed in string theory \cite{9711200}. Using this duality, the quantum properties in a conformal field theory can be investigated by studying its dual string theory in an AdS background. Lately, AdS/CFT duality has been generalized and applied to non-conformal field theory including QCD, namely AdS/QCD or holographic QCD \cite{0304032,0306018,0311270,0412141,0507073,0501128,0602229,0611099,0801.4383,0804.0434,0806.3830,1005.4690,1006.5461,1012.1864,1103.5389,1108.2029,1108.0684,1111.4953,1209.4512,1301.0385,1406.1865,1506.05930}. Holographic QCD offers a suitable frame to study phase transition in QCD at all temperature and chemical potential. There are two type of holographic QCD models, i.e. top-down and bottom-up models. In this work, we will study a bottom-up model based on the soft-wall model \cite{0602229}, which is the first holographic QCD model realizing the linear Regge spectrum of mesons\cite{0507246}. The holographic QCD model to describe heavy quarks has been constructed in \cite{1301.0385}. By analytic solving the full backreacted Einstein equation, meson Regge spectrum and phase structure in QCD were studied. In \cite{9803135,9803137,0604204,0610135,0611304,0701157,0807.4747,1004.1880,1008.3116,1201.0820,1206.2824,1401.3635}, a probe open string in an AdS background was studied. In \cite{1506.05930}, probe open strings were added in the holographic QCD model for heavy quarks and the behavior of open strings has been studied in the black hole background. By combining the various open strings configurations and phase structure of black hole background, a new physical picture of phase diagram for confinement-deconfinement phase transition in holographic QCD was suggested. The holographic QCD model to describe light quarks has also been tried in \cite{1406.1865}. However, the scalar field becomes complex causes some problems in \cite{1406.1865}.

\begin{figure}[t!]
\begin{center}
\includegraphics[
height=1.4in, width=5.3in]
{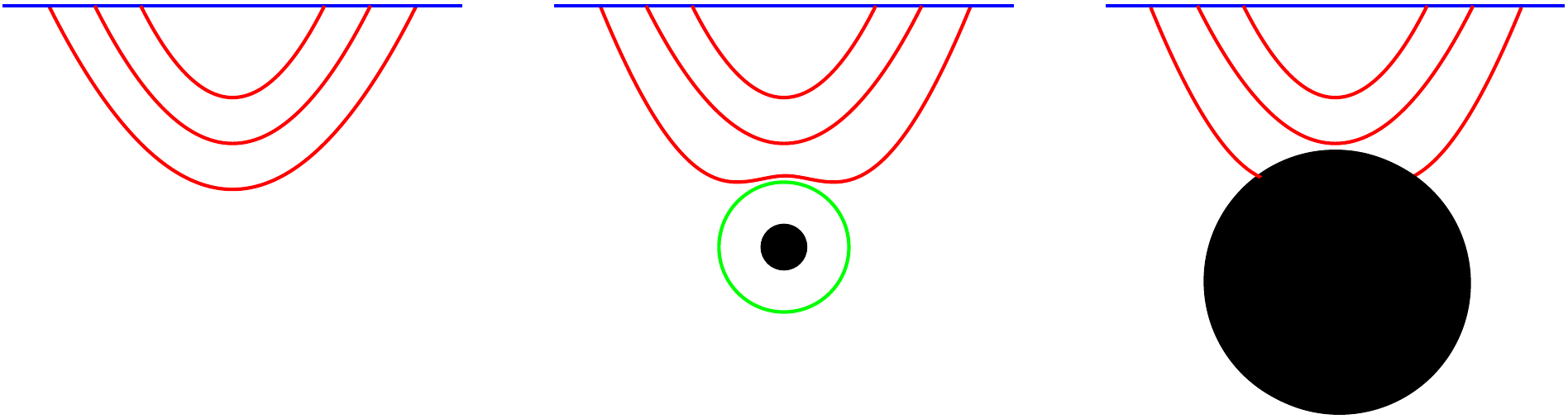}
\vskip 0.3 cm \hskip -0.0 cm (a) \hskip 4.0 cm (b) \hskip 4.0 cm (c)
\end{center}
\caption{Three configurations for open strings in black hole background. When black hole absent as (a), open strings are always connected with two ends attached on the AdS boundary. In the small black hole case (b), the strings can not exceed a certain depth from the boundary and still connected with their two ends on the AdS boundary. For the large black hole case (c), the open string with their two ends separated enough will break into two open strings where one end stick on the AdS boundary another falls in the black hole horizon.} \label{fig_BHstring}
\end{figure}

In this work, we construct a bottom-up holographic QCD model by studying its dual 5-dimensional gravity theory coupled to a Abelian gauge field and a neutral scalar field, i.e. Einstein-Maxwell-scalar system (EMS) . We analytical solve the equations of motion to obtain a family of black hole backgrounds which depend on two arbitrary functions $f(z)$ and $A(z)$. Since one of the crucial properties for the soft-wall holographic QCD models is that the vector meson spectrum satisfies the linear Regge trajectories at zero temperature and zero density, we are able to fix the function $f(z)$ by requiring the linear meson spectrum. Then, by choosing a suitable function $A(z)$, we obtain a black hole background which appropriately describe many important properties in QCD. We explore the phase structure of the black hole background by studying its thermodynamically quantities under different temperature and chemical potential. In addition, we study the Wilson loop \cite{9803002} and the light quark potential in our holographic QCD model by putting a probe open string in the black hole background and studying the dynamics of the open string. We found three configurations for the open string in a black hole as in Fig.\ref{fig_BHstring}. Combining the background phase structure and the open string breaking effect, we obtain the phase diagram for confinement-deconfinement transition.

The paper is organized as follows. In section II, we review the EMS system and obtain a family of analytic black hole background solutions. We study the thermodynamics of the black hole backgrounds to get their phase structure in section III. We add probe open string in the black hole background to study their various configurations. We calculate the expectation value of the Wilson loop and study the quark potential in section IV. By combining the background phase structure and the open string breaking effect, we obtain the phase diagram for confinement-deconfinement transition in section V. We conclude our result in section VI.

\setcounter{equation}{0}
\renewcommand{\theequation}{\arabic{section}.\arabic{equation}}%

\section{Einstein-Maxwell-Scalar System}

The EMS system has been widely studied in constructing bottom-up holographic QCD models. In this section, we briefly review EMS and, by solving the equations of motion analytically, we obtain a family of black hole solutions which the authors have previously studied in \cite{1301.0385,1406.1865,1506.05930}.

\subsection{String Frame and Einstein Frame}

We consider a 5-dimensional EMS system with probe matter fields. The system can be described by an action with two parts, background sector $S_{B}$ and matter sector $S_{m}$, respectively,
\begin{equation}
S=S_{B}+S_{m}. \label{eq_S}
\end{equation}
In the string frame, labeled by a sup-index $s$, the background part $S_{B}$ includes a gravity field $g^s_{\mu\nu}$, a Maxwell field $A_{\mu}$ and a neutral scalar field $\phi^s$,
\begin{equation}
S_{B} = \frac{1}{16\pi G_{5}} \int d^{5}x\sqrt{-g^s}e^{-2\phi^s}
	\left[{R^s-\frac{f^s_B(\phi^s)}{4}{F}^{2}}
			+4\partial_{\mu}\phi^s\partial^{\mu}\phi^s
			-V^s(\phi^s)  \right], \label{eq_SB_sf}
\end{equation}
where ${G}_{5}$ is the 5-dimensional Newtonian constant, ${{F}}_{\mu\nu} = \partial_{\mu}{A}_{\nu}-\partial_{\nu}{A}_{\mu}$ is the gauge field strength corresponding to the Maxwell field, $f^s_B(\phi^s)$ is the gauge kinetic function associated to Maxwell field and $V^s(\phi^s)$ is the potential of the scalar field.
The matter part $S_{m}$ of the MES system includes massless gauge fields $A_{\mu}^{V}$ and $A_{\mu}^{\tilde{V}}$, which are treated as probes in the background, describing the degrees of freedom of vector mesons and pseudovector mesons on the 4-dimensional boundary,
\begin{equation}
S_{m} = -\frac{1}{16\pi G_{5}}\int d^{5}x\sqrt{-g^s}e^{-2\phi^s}
		\left[{\frac{f^s_{m}\left(\phi^s\right)}{4}}
		\left(F_{V}^{2}+F_{\tilde{V}}^2\right)\right], \label{eq_Sm_sf}
\end{equation}
where $f^s_m(\phi^s)$ is the gauge kinetic function of the gauge fields $A_{\mu}^{V}$ and $A_{\mu}^{\tilde{V}}$. It is worth to mention that the gauge kinetic functions $f^s_B$ and $f^s_m$ are positive-defined and are not necessary to be the same. For simplicity, in this paper, we set $f^s_B=f^s_m=f^{s}$.
We have constructed the EMS system in the string frame, in which it is natural to impose the physical boundary conditions when solving the background solution according to the AdS/CFT dictionary. However, it is more convenient to solve the equations of motion and study the thermodynamical properties of QCD in Einstein frame.

The string frame action is characterized by the exponential dilaton factor in front of the Einstein term, i.e. $e^{-2\phi^s}R^s$. To transform the action from string frame to Einstein frame, in which the Einstein term is expressed in the conventional form, we make the following Weyl transformations,
\begin{equation} \label{eq_weyl trans}
\phi^{s}=\sqrt{\frac{3}{8}}\phi ,~
g^s_{\mu\nu}=g_{\mu\nu} {e}^{\sqrt{\frac{2}{3}}\phi} ,~
f^{s}\left(\phi^{s}\right)=f\left(\phi\right) {e}^{\sqrt{\frac{2}{3}}\phi} ,~
V^{s}\left(\phi^{s}\right)={e}^{-\sqrt{\frac{2}{3}}\phi}V\left(\phi\right).
\end{equation}
Thus, in Einstein frame, the actions in Eqs.(\ref{eq_SB_sf}-\ref{eq_Sm_sf}) become
\begin{eqnarray}
S_{B} &=& \frac{1}{16\pi G_{5}} \int d^{5}x\sqrt{-g}
	\left[{R-\frac{f\left(\phi\right)}{4}F^{2}}
			-\frac{1}{2}\partial_{\mu}\phi \partial^{\mu}\phi
			-V\left(\phi\right) \right], \label{eq_SB_Ef} \\
S_{m} &=& -\frac{1}{16\pi G_{5}}\int d^{5}x\sqrt{-g}
		\left[{\frac{f\left(\phi\right)}{4}}
		\left(F_{V}^{2}+F_{\tilde{V}}^2\right)\right]. \label{eq_Sm_Ef}
\end{eqnarray}

\subsection{Black Holes Solution}
Now we are in the stage to derive the equations of motion of our EMS system from Eqs.(\ref{eq_SB_Ef}-\ref{eq_Sm_Ef}). We first study the background by turning off the probe matters of vector field $A_{\mu}^{V}$ and pseudovector field $A_{\mu}^{\tilde{V}}$, the equations of motion can be derived as
\begin{eqnarray}
\nabla^{2}\phi &=& \frac{\partial V}{\partial\phi}+\frac{F^2}{4}\frac{\partial f}{\partial\phi}, \label{eq_eom_phi}\\
\nabla_{\mu}\left[ f(\phi)F^{\mu\nu} \right] &=&0, \label{eq_eom_A}\\
R_{\mu\nu}-\frac{1}{2} g_{\mu\nu}R &=& \frac{f(\phi)}{2} \left( F_{\mu\rho}F_{\nu}^{~\rho}-\frac{1}{4}g_{\mu\nu}F^{2}\right) +\frac{1}{2}\left[\partial_{\mu}\phi\partial_{\nu}\phi-\frac{1}{2}g_{\mu\nu}\left(\partial\phi\right)^{2}-g_{\mu\nu}V(\phi)\right] \label{eq_eom_g}.
\end{eqnarray}
Since we are going to study the thermodynamical properties of QCD at finite temperature by gauge/gravity correspondence, we consider the following blackening ansatz of the background metric in Einstein frame as
\begin{eqnarray}
ds^{2} &=& \frac{e^{2A\left(z\right)}}{z^{2}}
			\left[-g(z)dt^{2} +d\vec{x}^{2} +\frac{dz^{2}}{g(z)}\right],\label{eq_metric}\\
\phi &=& \phi\left(z\right),~ A_{\mu}=\left(A_{t}\left(z\right),\vec{0},0\right), \label{eq_ansatz}
\end{eqnarray}
where $z = 0$ corresponds to the conformal boundary of the 5-dimensional space-time and $g(z)$ stands for the blackening factor. Here we have set the radial of $AdS_5$ to be unit by scale invariant.

Plugging the ansatz in Eq.(\ref{eq_metric}-\ref{eq_ansatz}) into the equations of motion (\ref{eq_eom_phi}-\ref{eq_eom_g}) leads to the following equations of motion for the background fields,
\begin{eqnarray}
\phi^{\prime\prime}+\left(\frac{g^{\prime}}{g}+3A^{\prime}-\dfrac{3}{z}\right) \phi^{\prime}+\left( \frac{z^{2}e^{-2A}A_{t}^{\prime2}f_{\phi}}{2g}-\frac{e^{2A}V_{\phi}}{z^{2}g}\right)   &=&0,\label{eom_phi}\\
A_{t}^{\prime\prime}+\left(  \frac{f^{\prime}}{f}+A^{\prime}-\dfrac{1}{z}\right)  A_{t}^{\prime}  &=&0,\label{eom_At}\\
A^{\prime\prime}-A^{\prime2}+\dfrac{2}{z}A^{\prime}+\dfrac{\phi^{\prime2}}{6}&=& 0,\label{eom_A}\\
g^{\prime\prime}+\left(  3A^{\prime}-\dfrac{3}{z}\right)  g^{\prime}-e^{-2A}z^{2}fA_{t}^{\prime2}  &=& 0,\label{eom_g}\\
A^{\prime\prime}+3A^{\prime2}+\left(  \dfrac{3g^{\prime}}{2g}-\dfrac{6}{z}\right)  A^{\prime}-\dfrac{1}{z}\left(  \dfrac{3g^{\prime}}{2g}-\dfrac{4}{z}\right)  +\dfrac{g^{\prime\prime}}{6g}+\frac{e^{2A}V}{3z^{2}g}  &=& 0.
\label{eom_V}
\end{eqnarray}
Next, we specify the physical boundary conditions to solve the Eqs.(\ref{eom_phi}-\ref{eom_V}). We impose the conditions that the metric in the string frame to be asymptotic to $AdS_5$ at the boundary $z=0$ and the black hole solutions are regular at the horizon $z=z_H$.
\begin{enumerate}[(i)]
    \item $z \to 0:$
		\begin{equation}
 		A(0)+\sqrt{\frac{1}{6}}\phi(0)=0,~ g(0)=1. \label{bdy_0}
		\end{equation}
    \item $z=z_H:$
        \begin{equation}
 		A_t(z_H)=g(z_H)=0. \label{bdy_zh}
		\end{equation}
\end{enumerate}
It is natural to introduce concepts of chemical potential $\mu$ and baryon density $\rho$ in QCD from the temporal component of gauge field $A_t$ by the holographic dictionary of the gauge/gravity correspondence,
\begin{equation}
A_t(z)=\mu+\rho z^2 + O(z^4). \label{eq_At}
\end{equation}
As mentioned in the introduction, one of the crucial properties for the soft-wall holographic QCD models is that the vector meson spectrum satisfies the linear Regge trajectories at zero temperature and zero density, i.e. $\mu=\rho=0$. This issue was first
addressed in the soft-wall model \cite{0602229} using the method of AdS/QCD duality.

We consider the 5-dimensional probe vector field $V$ in the action (\ref{eq_Sm_sf}). The equation of motion for the vector field reads
\begin{equation}
\nabla_{\mu}\left[  f\left(  \phi\right)  F_{V}^{\mu\nu}\right]  ={{0.}}%
\end{equation}
Following \cite{0602229}, we first use the gauge invariance to fix the gauge
$V_{z}=0$, then the equation of motion of the transverse vector field $V_{\mu
}$ $\left(  \partial^{\mu}V_{\mu}=0\right)  $ in the background (\ref{eq_metric})
reduces to%
\begin{equation}
-\psi_{i}^{\prime\prime}+U\left(  z\right)  \psi_{i}=\left(  \dfrac{\omega
^{2}}{g^{2}}-\dfrac{p^{2}}{g}\right)  \psi_{i}, \label{eqv}%
\end{equation}
where we have performed the Fourier transformation for the vector field
$V_{i}$ as%
\begin{equation}
V_{i}\left(  x,z\right)  =\int\dfrac{d^{4}k}{\left(  2\pi\right)  ^{4}%
}e^{ik\cdot x}v_{i}\left(  z\right)  , \label{V-v}%
\end{equation}
and further redefined the functions $v_{i}\left(  z\right)  $ with%
\begin{equation}
v_{i}=\left(  \dfrac{z}{e^{A}fg}\right)  ^{1/2}\psi_{i}\equiv X\psi_{i},
\end{equation}
with the potential function%
\begin{equation}
U\left(  z\right)  =\dfrac{2X^{\prime2}}{X^{2}}-\dfrac{X^{\prime\prime}}{X}.
\end{equation}
In the case of zero temperature and zero chemical potential, we expect that the discrete spectrum of the vector mesons obeys the linear Regge trajectories. The above Eq.(\ref{eqv}) reduces to a Schr\"{o}dinger
equation%
\begin{equation}
-\psi_{i}^{\prime\prime}+U\left(  z\right)  \psi_{i}=m^{2}\psi_{i},
\label{Schrodinger}%
\end{equation}
where $-m^{2}=k^{2}=-\omega^{2}+p^{2}$. To produce the discrete mass spectrum with the linear Regge trajectories, the potential $U\left( z \right)$ should be in certain forms. A simple choice is to fix the gauge kinetic function as
\begin{equation}
f\left(  z\right)  =e^{\pm cz^{2}-A\left(  z\right)  },
\end{equation}
which leads the potential to be
\begin{equation}
U\left(  z\right)  =-\dfrac{3}{4z^{2}}-c^{2}z^{2}. \label{potential}
\end{equation}
The Schr\"{o}dinger Eq.(\ref{Schrodinger}) with the potential in
Eq.(\ref{potential}) has the discrete eigenvalues
\begin{equation}
m_{n}^{2}=4cn, \label{mass}%
\end{equation}
which is linear in the energy level $n$ as we expect for the vector spectrum at zero temperature and zero density which was well known as the linear Regge trajectories \cite{0507246}.

\begin{figure}[t!]
\begin{center}
\includegraphics[
height=2.2in, width=3.2in]
{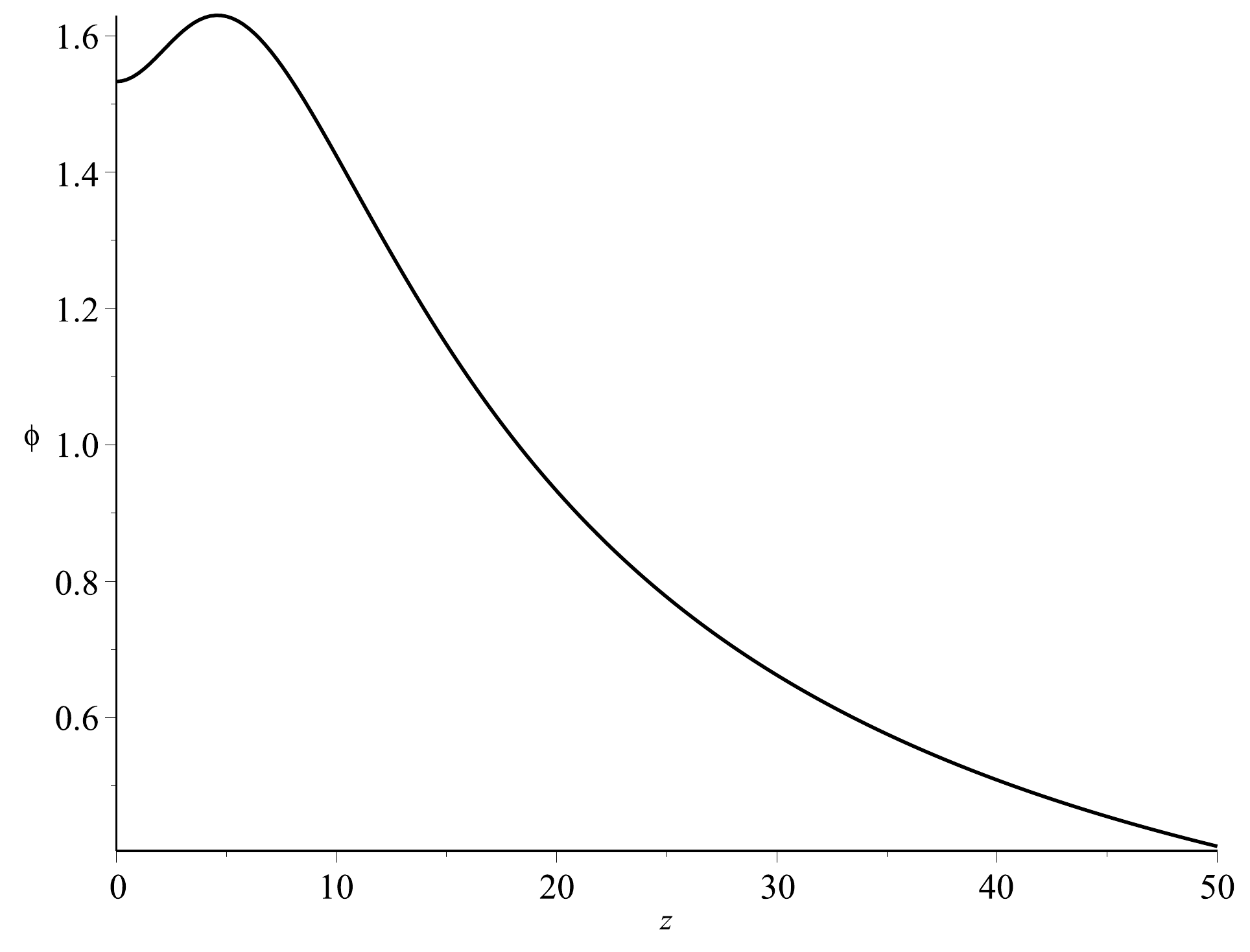}
\includegraphics[
height=2.2in, width=3.2in]
{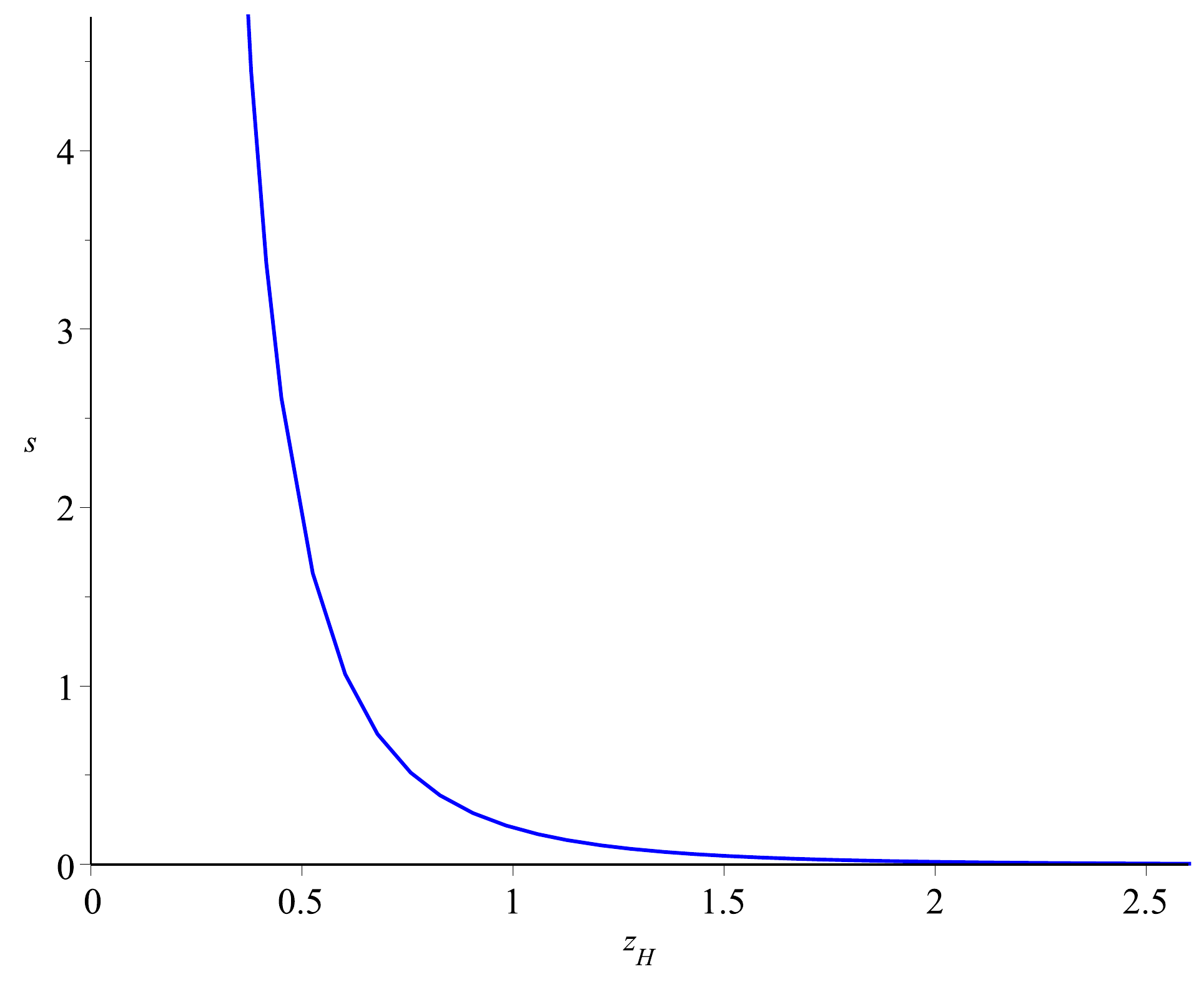}
\vskip -0.05cm \hskip 0.15 cm (a) \hskip 5.5 cm (b)
\end{center}
\caption{(a) The neutral scalar field $\phi$ is a positive well defined function in z for suitable choice of the parameters of warped factor $A$. (b) The black hole entropy is monotonous decrease as horizon increasing.} \label{fig_phiz} \label{fig_szh}
\end{figure}

Once we fix the gauge kinetic function $f(z)$, the equations of motion (\ref{eom_At}-\ref{eom_V}) can be analytically solved as
\begin{eqnarray}
\phi^{\prime}\left(  z\right) &=&\sqrt{-6\left( A^{\prime\prime}
-A^{\prime2}+\dfrac{2}{z}A^{\prime}\right)  },\label{phip-A}\\
A_{t}\left(  z\right)   &=& \mu\dfrac{e^{cz^{2}}-e^{cz_{H}^{2}}}%
{1-e^{cz_{H}^{2}}},\label{At-A}\\
g\left(  z\right) &=& 1-\dfrac{1}{\int_{0}^{z_{H}}y^{3}e^{-3A}dy}
\left[ \int_{0}^{z}y^{3}e^{-3A}dy
-\dfrac{2c\mu^{2}}{\left(  1-e^{cz_{H}^{2}}\right)  ^{2}}\left\vert
\begin{array}
[c]{cc}%
\int_{0}^{z_{H}}y^{3}e^{-3A}dy & \int_{0}^{z_{H}}y^{3}e^{-3A}e^{cy^{2}}dy\\
\int_{z_{H}}^{z}y^{3}e^{-3A}dy & \int_{z_{H}}^{z}y^{3}e^{-3A}e^{cy^{2}}dy
\end{array}
\right\vert \right]  ,\\
V\left(  z\right) &=&-3z^{2}ge^{-2A}\left[  A^{\prime\prime}+3A^{\prime
2}+\left(  \dfrac{3g^{\prime}}{2g}-\dfrac{6}{z}\right)  A^{\prime}-\dfrac
{1}{z}\left(  \dfrac{3g^{\prime}}{2g}-\dfrac{4}{z}\right)  +\dfrac
{g^{\prime\prime}}{6g}\right] . \label{V-A}
\end{eqnarray}
Eqs.(\ref{phip-A}-\ref{V-A}) represent a family of solutions for the black hole background depending on the warped factor $A\left(z\right)$, which could be arbitrary function which satisfies the boundary condition in Eq.(\ref{bdy_0}). Furthermore, we also need to ensure that the expression under the squared root in Eq.(\ref{phip-A}) is positive for $z\in(z,z_H)$ to guarantee a real scalar field $\phi(z)$. A simple choice of $A\left(z\right)=az^2+bz^4$ with $a,b<0$ has been used to study the phase structure of heavy quarks in holographic QCD in \cite{1301.0385}. In \cite{1406.1865}, a similar form has been tried to study the phase structure of light quarks in holographic QCD, but with a problem that the scalar field $\phi(z)$ becomes complex for some $z$. In this work, we choose the warped factor $A(z)$ as
\begin{equation}
A \left( z \right)= -a \ln (bz^2+1). \label{eq-Ae}
\end{equation}
Plugging Eq.(\ref{eq-Ae}) into Eq.(\ref{phip-A}), it is easy to show that the scalar field $\phi(z)$ is always real for positive $a$ and $b$ as shown in Fig.\ref{fig_phiz}(a). By fitting the mass spectrum of $\rho$ meson with its excitations and comparing the phase transition temperature with lattice calculation, we can determine the parameters in Eq.(\ref{potential}) and Eq.(\ref{eq-Ae}) as $a=4.046, ~b=0.01613$ and $c=0.227$.

\setcounter{equation}{0}
\renewcommand{\theequation}{\arabic{section}.\arabic{equation}}
\section{Phase Structure of the Black Hole Background}
In this section, we will explore the phase structure of the black hole background in Eqs.(\ref{phip-A}-\ref{V-A}) which we obtained in the last section. The entropy density and the Hawking temperature can be calculated as,
\begin{equation}
s=\frac{\sqrt{g(\vec{x})}}{4}\bigg\vert_{z_H}, ~T=\frac{g'(z_H)}{4\pi}, \label{eq_def_sT}\\
\end{equation}
where $g(\vec{x})$ represents the metric of the internal space along $\vec{x}$.
The free energy in grand canonical ensemble can be obtained from the first law of thermodynamics ,
\begin{equation}
F=\epsilon-Ts-\mu\rho, \label{eq_def_F}
\end{equation}
where $\epsilon$ labels the internal energy density. Comparing the free energies between different sizes of black holes at the same temperature under certain finite value of chemical potential, we are able to obtain the phase structure of black holes which corresponds to the phase structure in the holographic QCD due to AdS/CFT correspondence.

\subsection{Black hole thermodynamics}
The entropy density, defined in Eq.(\ref{eq_def_sT}), can be easily obtained for our black hole background in Eqs.(\ref{phip-A}-\ref{V-A}) as,
\begin{equation}
s=\frac{e^{3A(z_H)}}{4z_H^{3}},
\end{equation}
which is plotted in Fig.\ref{fig_szh}(b). We see that the entropy density is a monotonously decreasing function as horizon increasing. Based on the second law of thermodynamics, it implies that our black hole background prefers smaller size with larger entropy. It is worth to mention, the smaller value of horizon relates the bigger size of black hole.

\begin{figure}[t!]
\begin{center}
\includegraphics[
height=2.2in, width=3.2in]
{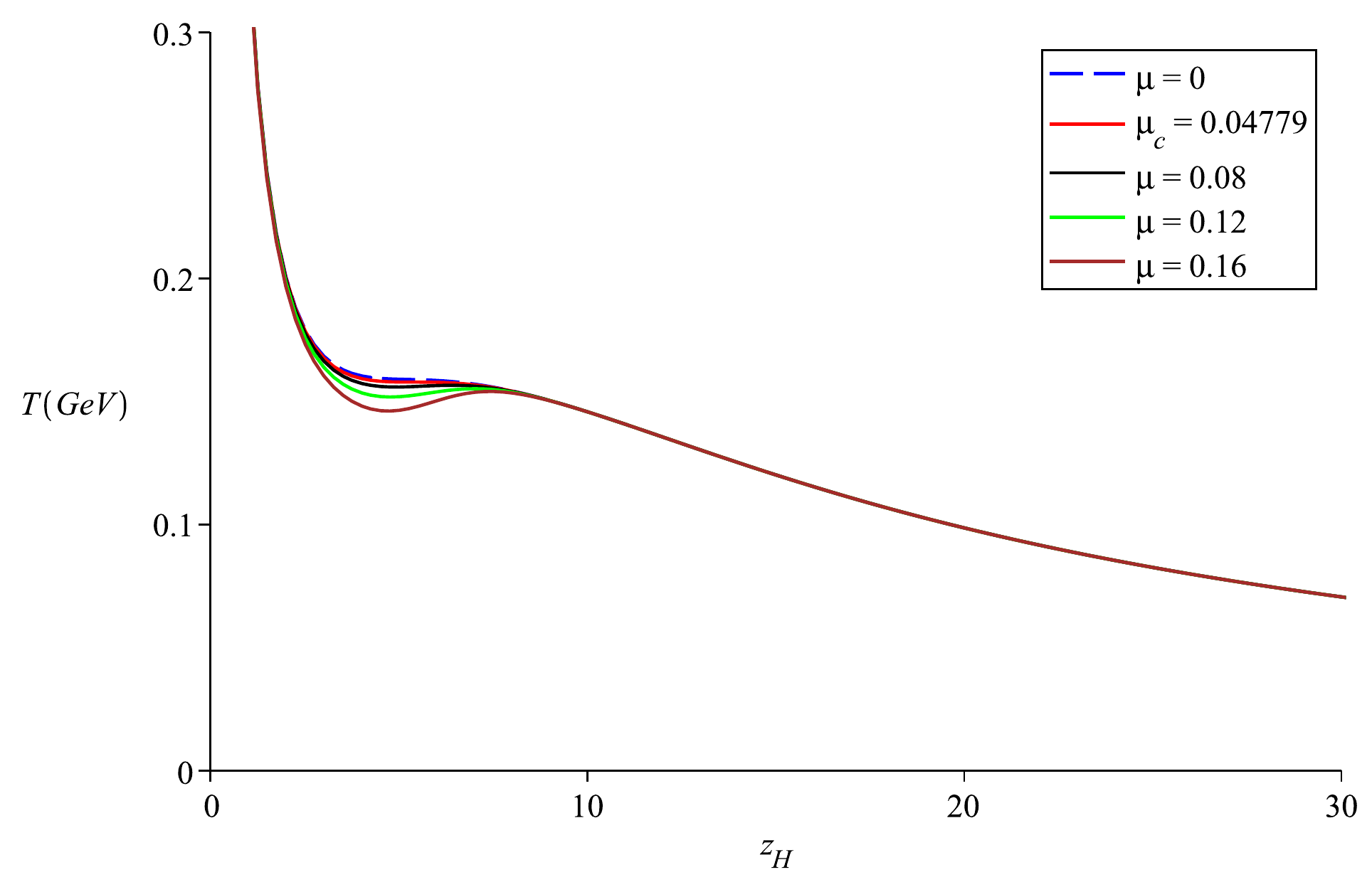}
\includegraphics[
height=2.2in, width=3.2in]
{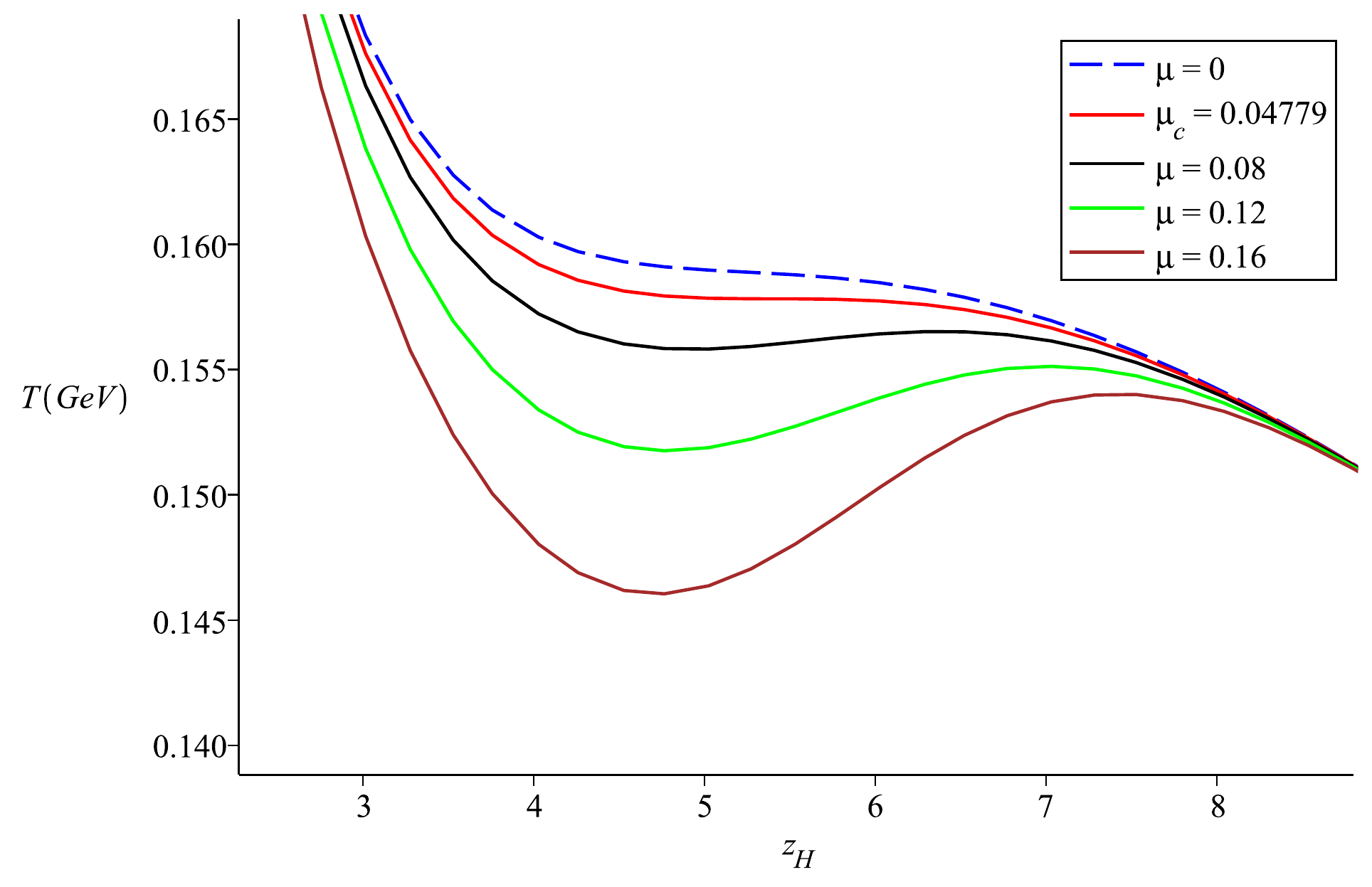}
\vskip -0.05cm \hskip 0.15 cm (a) \hskip 5.5 cm (b)
\end{center}
\caption{(a) The black hole temperature v.s. horizon at different chemical potentials, where $\mu_c=0.04779~GeV$. We enlarge the phase transition region in (b) to display the detail structure.} \label{fig_T}
\end{figure}

Another crucial thermal quantity in studying phase transition is the black hole temperature, also defined in Eq.(\ref{eq_def_sT}), which can be calculated for our black hole background as
\begin{equation}
T=\dfrac{z_{H}^{3}e^{-3A\left( z_{H} \right) }} {4\pi \int_{0}^{z_{H}} y^{3}e^{-3A}dy}
    \left[ 1-
    \dfrac{2c\mu^{2} \left( e^{cz_{H}^{2}}\int_{0}^{z_{H}}y^{3}e^{-3A}dy-\int_{0}^{z_{H}}y^{3}e^{-3A}e^{cy^{2}}dy \right) } {\left( 1-e^{cz_{H}^{2}} \right)^{2}}
    \right]  .
\end{equation}
The temperature as the function of the horizon, at various chemical potentials, are plotted in Fig.\ref{fig_T}. At small chemical potential, $0 \leq \mu < \mu_c$, the temperature behaves as a monotonous function of horizon and decreases to zero at infinity horizon. It is clear that there is no phase transition because of the monotonous behavior of temperature . At large chemical potential, $\mu \geq \mu_c$, the temperature becomes multivalued which implies that there will be a phase transition between black holes with different sizes.

\begin{figure}[t!]
\begin{center}
\includegraphics[
height=2.2in, width=3.2in]
{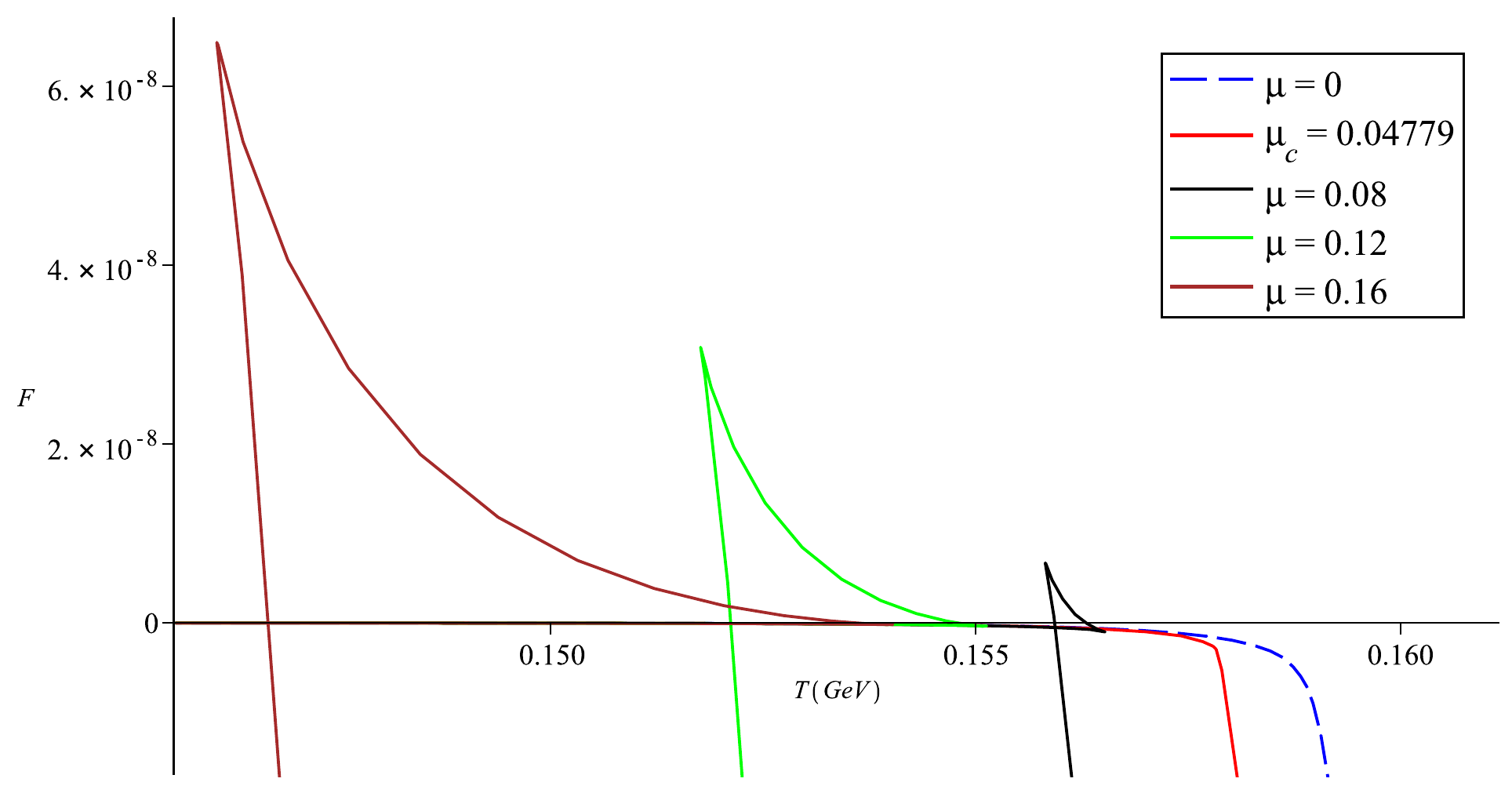}
\includegraphics[
height=2.2in, width=3.2in]
{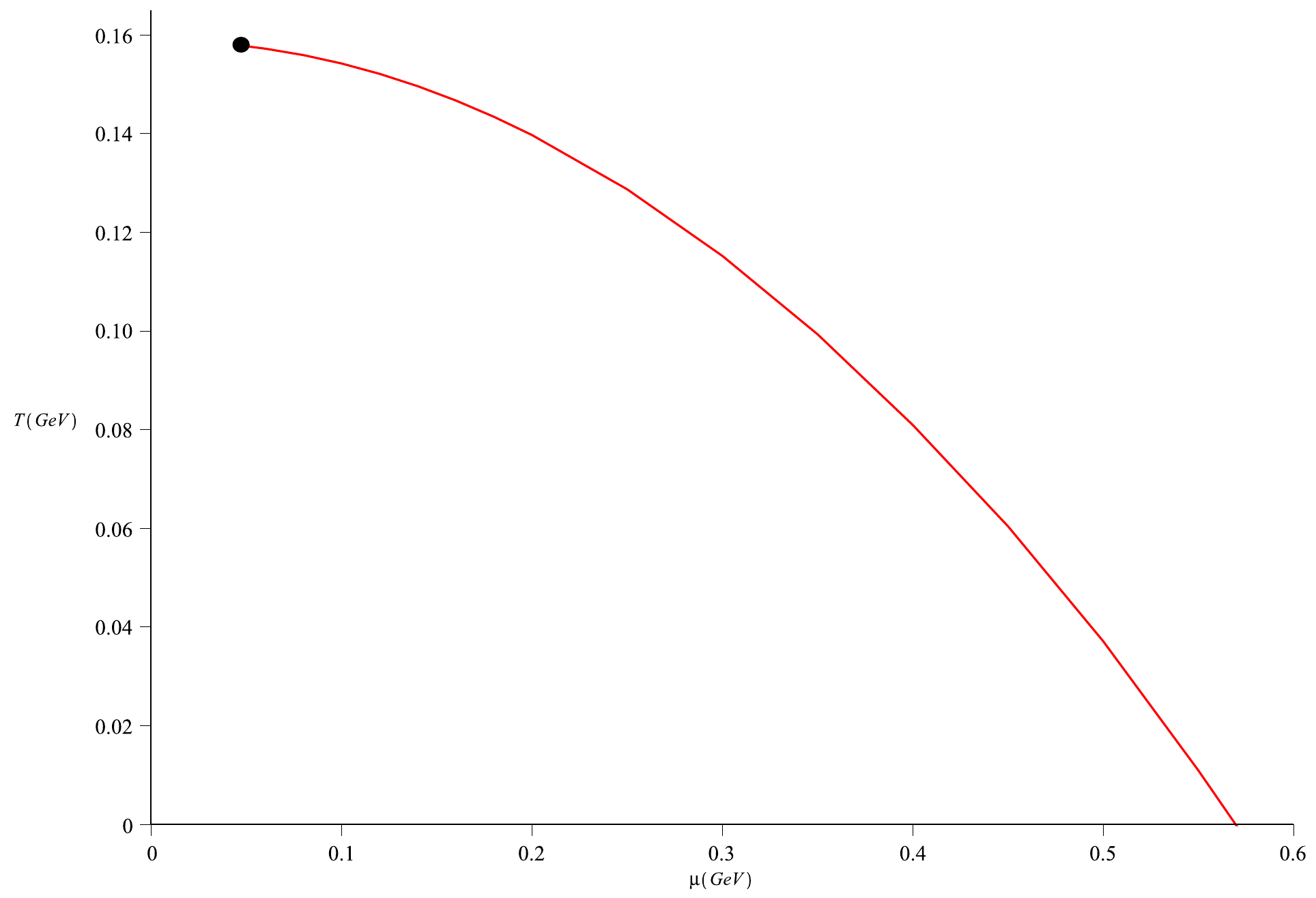}
\vskip -0.05cm \hskip 0.15 cm (c) \hskip 5.5 cm (d)
\end{center}
\caption{(a) The free energy v.s. temperature at various chemical potentials. The free energy behaves as a multivalued function of temperature with the swallow-tailed shapes at $\mu>\mu_c$ and becomes monotonous at $\mu<\mu_c$.
(b) The phase diagram in $T-\mu$ plane. For the large chemical potential $\mu>\mu_c$, the system undergoes a first order phase transition at a finite temperature and ends at the critical endpoint $(\mu_c,T_c) \simeq (0.04779,0.1578)$. For small chemical potential $0\leq \mu < \mu_c$, the phase transition reduces to a crossover. The zero temperature phase transition is located at $\mu_{T=0}=0.5695$.} \label{fig_FT} \label{fig_Tmu_BH}
\end{figure}

In order to determine the transition temperatures at each chemical potential $T_{BB}(\mu)$ between black holes with different sizes, we have to compute the free energy in grand canonical ensemble from the first law of thermodynamics. At fixed volume, we have
\begin{equation}
dF=-sdT-\rho d\mu.
\end{equation}
Thus, the free energy for a given chemical potential $\mu$ can be evaluated by an integral,
\begin{equation}
F= -\int sdT= \int_{zH}^\infty s(z_H)T'(z_H)dz_H, \label{eq_int_F}
\end{equation}
where we have normalized the free energy of the black holes vanishing at $z_H \to \infty$, i.e. $T=0$, which is equal to the free energy of the thermal gas. The free energy v.s. temperature at various chemical potentials is plotted in Fig.\ref{fig_FT}(a). The intersection of free energy implies that there exists a phase transition between two black holes with different sizes at the temperature $T=T_{BB}$.

For $\mu>\mu_c$, the free energy behaves as the swallow-tiled shape and shrinks into a singular point at $\mu=\mu_c$, then disappears for $\mu<\mu_c$. The behavior of the free energy implies that the system undergoes a first order phase transition at each fixed chemical potential $\mu>\mu_c$ and ends at the critical endpoint at $(\mu_c,T_c)$ where the phase transition becomes second order. For $\mu<\mu_c$, the phase transition reduces to a crossover. The phase diagram of black hole to black hole phase transition is plotted in Fig.\ref{fig_Tmu_BH}(b), which is consistent with the phase diagram for light quarks obtained in lattice QCD simulations \cite{1009.4089}.

\subsection{Susceptibility and equations of state}
\begin{figure}[t!]
\begin{center}
\includegraphics[
height=2.5in, width=3.6in]
{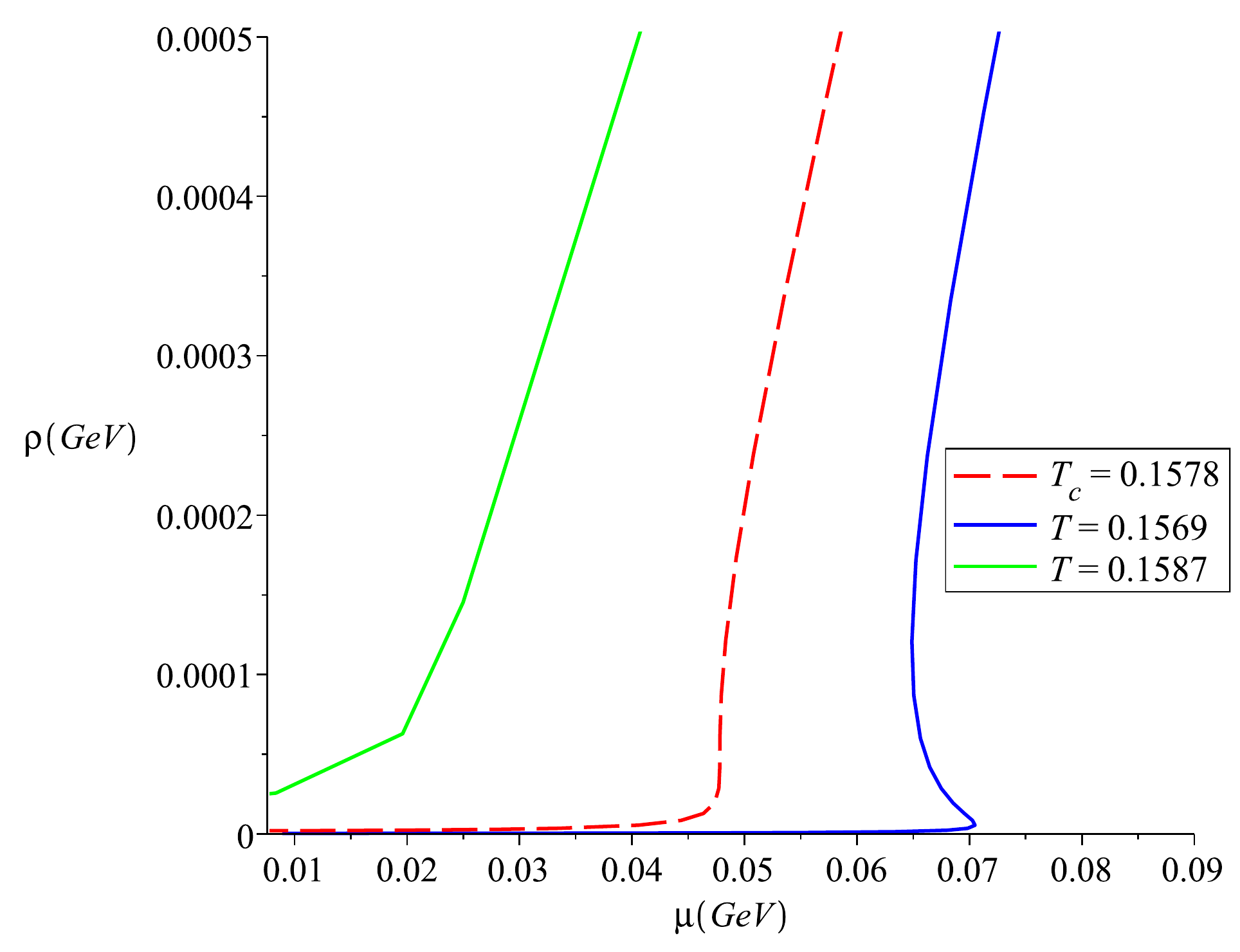}
\end{center}
\caption{The baryon density v.s. chemical potential. For $T < T_c$, $\rho$ is waving respect to $\mu$ until $T=T_c$ the waving pattern degenerates into a extremal profile then become monotonous for $T>T_c$.} \label{fig_rhomu}
\end{figure}

To justify the phase transition, we consider susceptibility and equations of state in the following. The susceptibility is defined as
\begin{equation}
\chi=\left(\frac{\partial \rho}{\partial \mu} \right)_T.
\end{equation}
We plot the baryon density $\rho$ v.s. chemical potential $\mu$ in Fig.\ref{fig_rhomu}. When $T < T_c$, the multivalued behavior indicates that there is a phase transition at certain values of chemical potential. While there is no transition happening for $T > T_c$, since $\rho$ is a single-valued function of $\mu$. At critical temperature $T_c$, the position where the slope becomes infinity locates critical point $(\mu_c,T_c)$ for the second phase transition, which is consistent with our previous result by comparing free energies between black holes with different sizes.	

\begin{figure}[t!]
\begin{center}
\includegraphics[
height=2.2in, width=3.2in]
{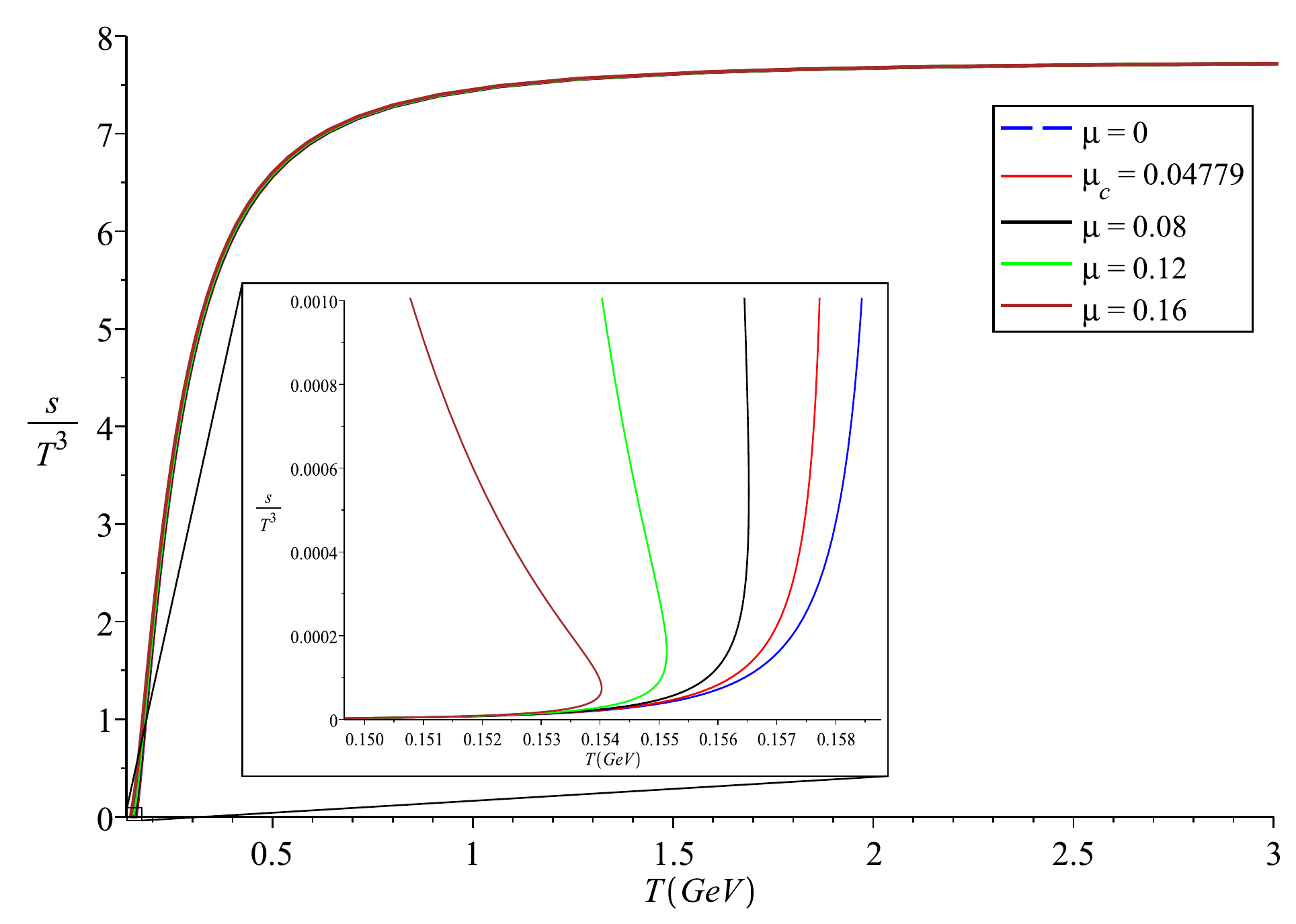}
\includegraphics[
height=2.2in, width=3.2in]
{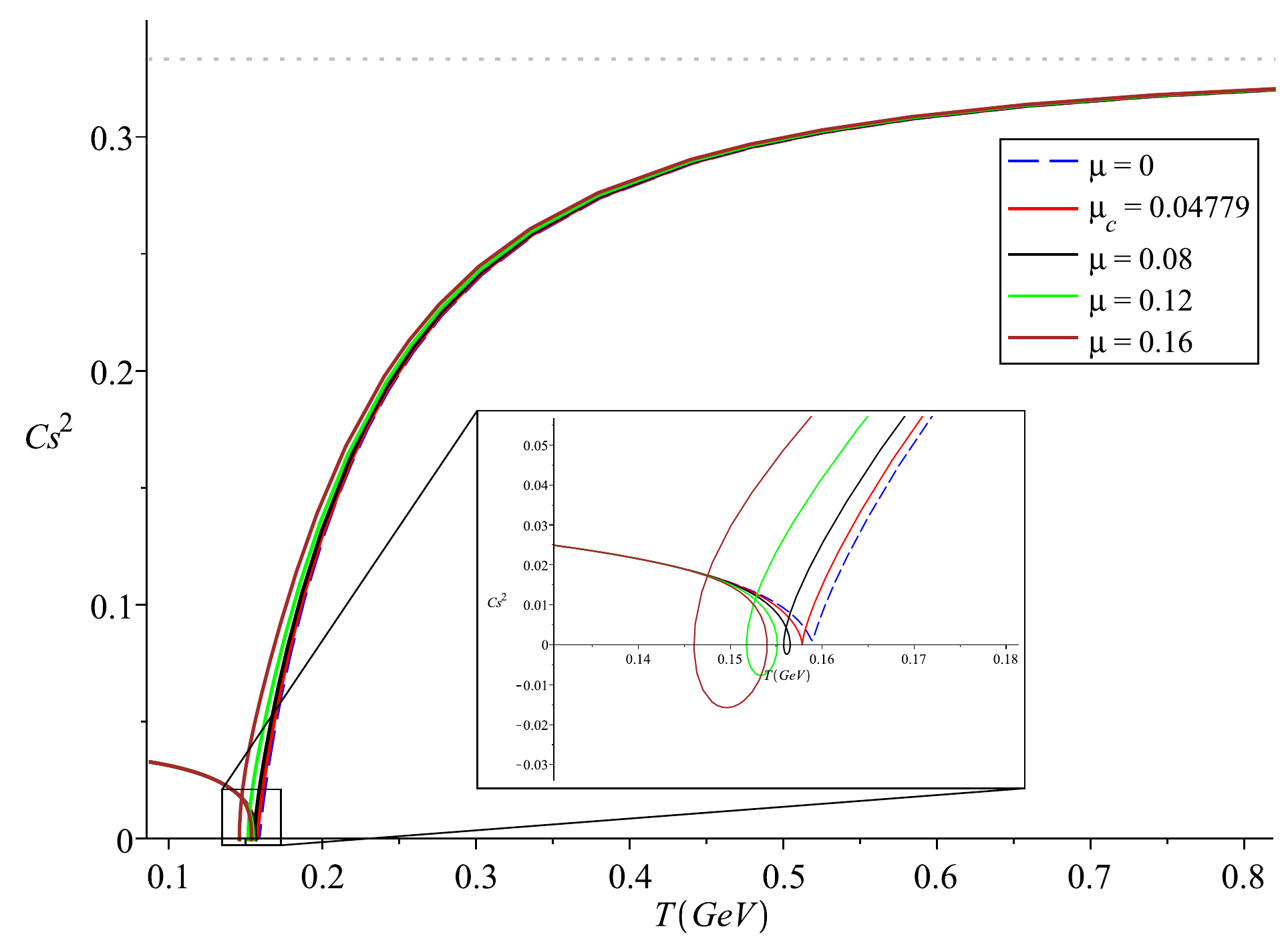}
\vskip -0.05cm \hskip 0.15 cm (a) \hskip 5.5 cm (b)\\
\includegraphics[
height=2.2in, width=3.2in]
{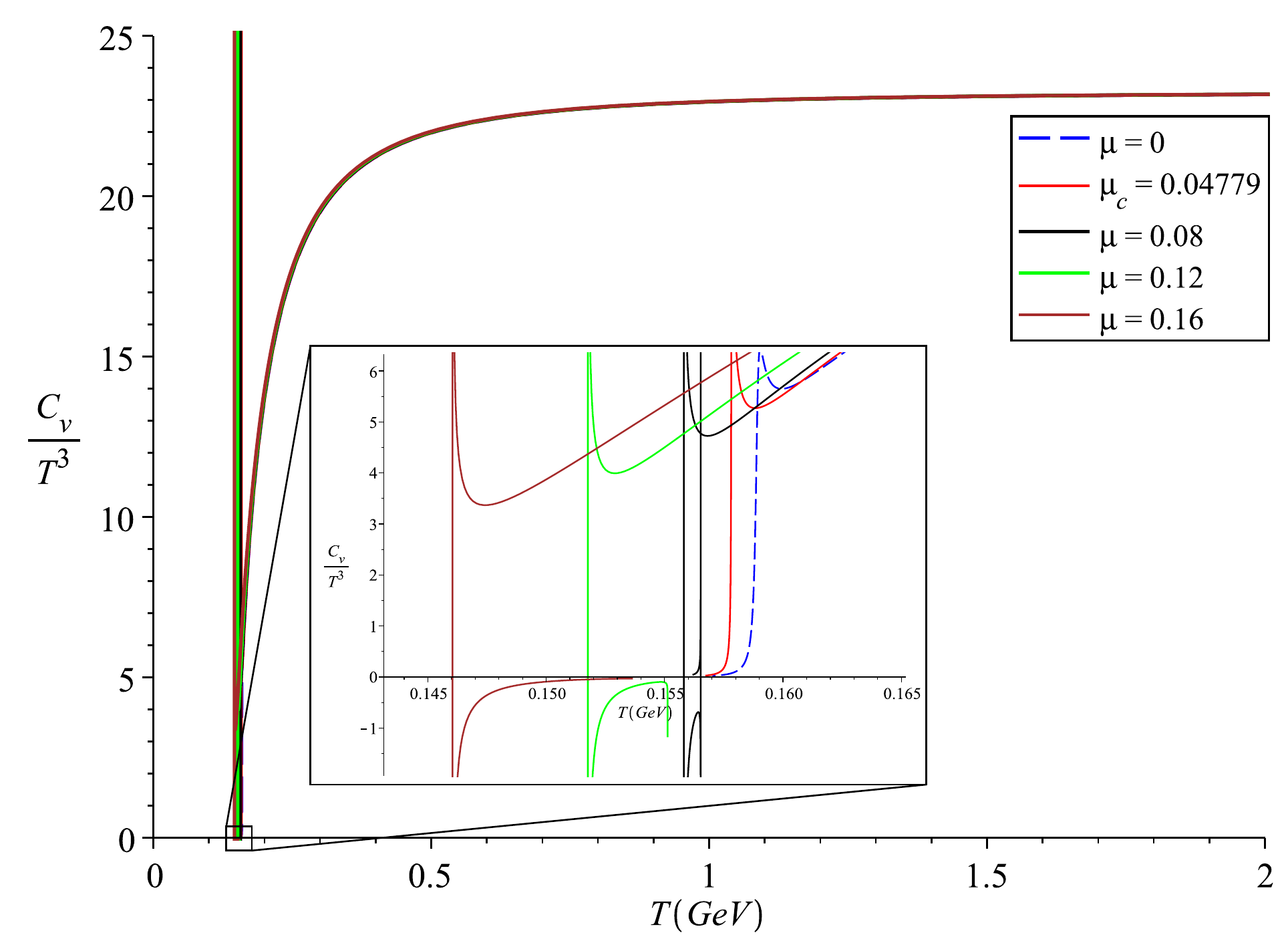}
\includegraphics[
height=2.2in, width=3.2in]
{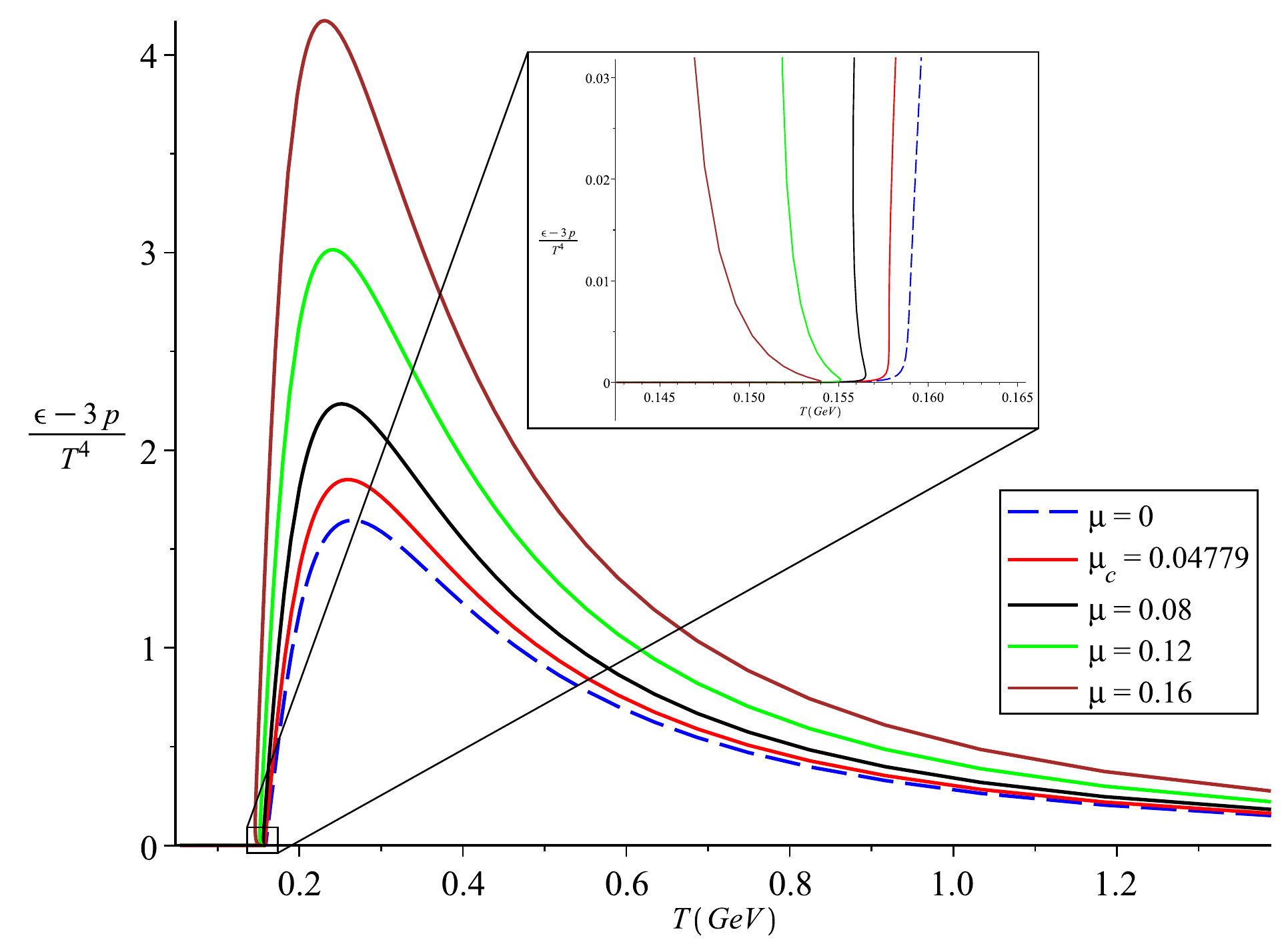}
\vskip -0.05cm \hskip 0.15 cm (c) \hskip 5.5 cm (d)
\end{center}
\caption{The normalized entropy (a), speed of sound (b), specific heat (c) and trace anomaly (d) v.s. temperature at various chemical potentials.} \label{fig_sT}\label{fig_csT} \label{fig_CvT} \label{fig_traceT}
\end{figure}

The normalized entropy density $s/T^3$ v.s. temperature $T$ is plotted in Fig.\ref{fig_sT}(a). The normalized entropy density becomes large in high temperature limit. The enlarged plot shows that the normalized entropy density is monotonous for $\mu<\mu_c$ and multivalued for $\mu>\mu_c$.

The square of speed of sound is defined as
\begin{equation}
c_s^2 = \frac{\partial\ln T}{\partial\ln s},
\end{equation}
which is plotted in Fig.\ref{fig_csT}(b). The positive/negative part of $c_s^2$ corresponds to the dynamical stable/unstable black hole. More precisely, the imaginary part of $c_s$ indicates the Gregory-Laflamme instability \cite{9301052, 9404071}, which is closely related to Gubser-Mitra conjecture that the dynamical stability of a horizon is equivalent to the thermodynamic stability \cite{0009126,0011127,0104071}. $c_s^2$ is always positive for $0 \leq \mu <\mu_c$, and reaches to zero at the critical  point $\mu_c, T_c$. It is worth to notice that $c_s^2$ approaches the conformal limit, $1/3$, in high temperature limit for every chemical potential $\mu$.
\begin{table}[t!]
\begin{center}
\begin{tabular}
[c]{|c|c|c|c|c|}\hline
Figure    & $\mu>\mu_c$  & $\mu=\mu_c$                    & $\mu<\mu_c$ \\\hline
$T-z_H$   & vibrating    & converges into a saddle point  & monotonous  \\\hline
$F-T$     & swallow-tail & gathers into a point           & monotonous  \\\hline
$c_s^2-T$ & knot         & shrinks into a cusp            & monotonous  \\\hline
$\rho-\mu$& waving       & narrows into an infinity slope & monotonous  \\\hline
$T-\mu$   & $1_{st}$ phase trans.   & $2_{nd}$ phase trans.          & crossover \\\hline
\end{tabular}
\end{center}
\caption{The significant patterns in the black hole phase transition.}
\label{table.BH}
\end{table}

One of the important quantities to realize the thermodynamically stability is specific heat capacity, which is defined as
\begin{equation}
C_v = T\left( \frac{\partial s}{\partial T} \right) = \frac{s}{c_s^2}.
\end{equation}
The normalized specific heat capacity $C_v/T^3$ v.s. temperature $T$ is plotted in Fig.\ref{fig_CvT}(c). For $0 \leq \mu < \mu_c$, $C_v$ is always positive indicating that the black holes are thermodynamically stable. On the other hand, as $\mu > \mu_c$, $C_v$ reveals the negative part which corresponds to the thermodynamically unstable. Furthermore, $C_v$ and $ c_s^2$ have exact the same behaves in sign. Therefore the imaginary part of speed of sound corresponds to the negative part of specific heat capacity, which implies that our system satisfies the Gubser-Mitra conjecture.

Trace anomaly $\epsilon-3p$ is another important thermodynamic quantity which can be derived from the internal energy,
\begin{equation}
\epsilon=F+Ts+\mu\rho.
\end{equation}
We plot the normalized trace anomaly $(\epsilon-3p)/T^4$ v.s. temperature $T$ in Fig.\ref{fig_traceT}(d). As the chemical potential decreasing, the peak of trace anomaly decreases and the multivalued behaviour becomes single-valued.

Finally, we summarize the behaviors of some important thermodynamic quantities in table.1.
\setcounter{equation}{0}
\renewcommand{\theequation}{\arabic{section}.\arabic{equation}}

\section{Open Strings in the Background}
\begin{figure}[t!]
\begin{center}
\includegraphics[
height=3.1in, width=2.7in]
{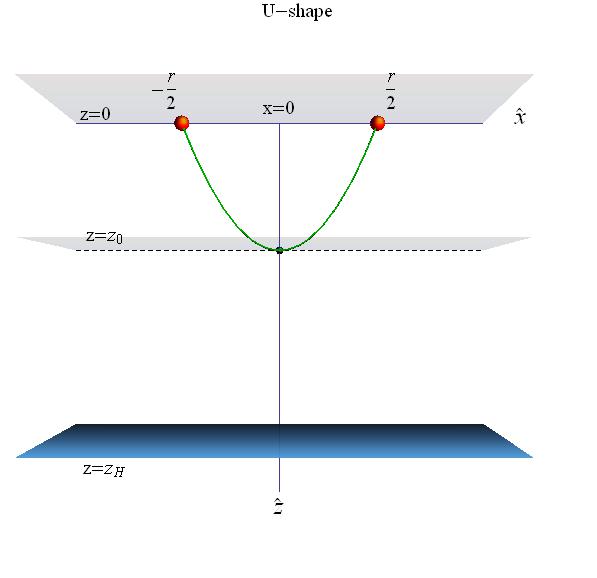}
\hspace*{0.5cm}
\includegraphics[
height=3.1in, width=2.7in]
{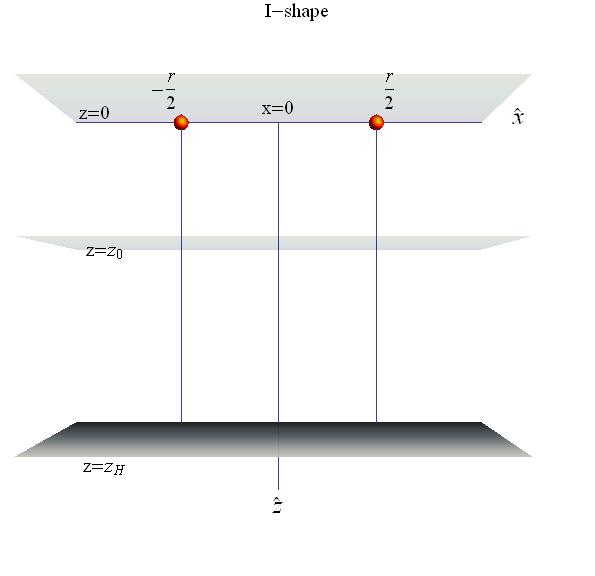}
\vskip -0.05cm \hskip 0.15 cm (a) \hskip 7.5 cm (b)
\end{center}
\caption{Two configurations of an open string in a black hole background. (a) The open string of U-shape configuration connects its two ends at the boundary $z=0$ and reaches the maximum depth at $z=z_0$. This configuration corresponds to a pair of quark-antiquark bounded state at the boundary. (b) The open string of I-shape configuration connects the boundary and the horizon at $z=z_H$. This configuration corresponds to the single quark/antiquark in the holographic QCD model.} \label{fig_UII_shape}
\end{figure}

In the following of this paper, we will study the phase structure for our holographic QCD model by adding probe open strings in the black hole backgrounds in Eqs.(\ref{phip-A}-\ref{V-A}). We consider open strings in the black hold background with their ends attaching on the boundary of the bulk at $z=0$ or black hole horizon at $z=z_H$. We find that there will be two configurations for an open string in the black hole background. One is the U-shape configuration with the open string reaching its maximum depth at $z=z_{0}$ and both of its ends attaching the boundary, another is the I-shape configuration with the straight open string having its two ends attached to the boundary and the horizon, respectively. The two configurations are showed in Fig.\ref{fig_UII_shape}.
Since the holographic QCD fields live on the boundary of the black hole background, it is natural to interpret the two ends of the open string as a quark-antiquark pair. The U-shape configuration corresponds to the quark-antiquark pair being connected by a string and can be identified as a meson state. While the I-shape configuration corresponds to a free quark or antiquark.

The Nambu-Goto action of an open string is
\begin{equation}
S_{NG}=\int d^{2}\xi\sqrt{-G}, \label{SNG}
\end{equation}
where the induced metric
\begin{equation}
G_{ab}=g_{\mu\nu}\partial_{a}X^{\mu}\partial_{b}X^{\nu}, \label{IM}
\end{equation}
on the 2-dimensional world-sheet that the string sweeps out as it moves with coordinates $(\xi^{0},\xi^{1})$ is the pullback of 5-dimensional target space-time metric $g_{\mu\nu_s}$,
\begin{equation}
ds^{2}=\frac{e^{2A_s(z)}}{z^{2}}
		\left( g(z)dt^{2}+d\vec{x}^{2}+\frac{1} {g(z)}dz^{2}\right),
\end{equation}
where we consider the Euclidean metric to study the thermal properties of the system to identify the black hole temperature of gravitational theory in bulk as thermal field theory on the boundary.

\subsection{Wilson Loop}
It is known that one can read off the energy of such a pair of quark-antiquark from the expectation value of the Wilson loop \cite{9803002,0604204,1201.0820},
\begin{equation}
\left\langle W \left( \mathcal{C} \right) \right\rangle
\sim e^{-V_{q\bar{q}}(r,T)/T}, \label{eq_WL1}
\end{equation}
where the rectangular Wilson loop $\mathcal{C}$ is along the directions $(t,x)$ on the boundary of the AdS space attached by a pair of the quark and antiquark separated by $r$, and $V(r,T)$ is the quark-antiquark potential.

\begin{figure}[t!]
\begin{center}
\includegraphics[
height=1.6in, width=3.6in]
{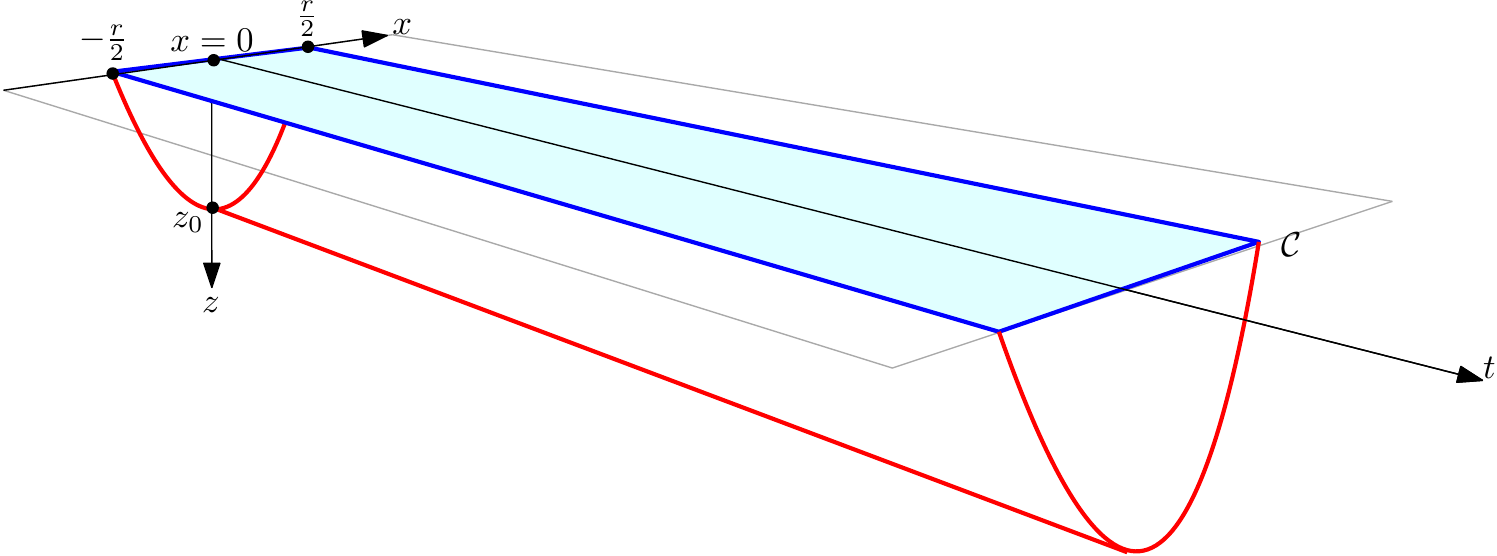}
\end{center}
\caption{The Wilson loop $\mathcal{C}$ could be viewed as the boundary of string world-sheet.} \label{fig_WL2}
\end{figure}

Based on string/gauge correspondence, if we consider a pair of quark-antiquark at $\left(z=0, x=\pm r/2\right)$ are connected by an open string as in Fig.\ref{fig_WL2}.
the expectation value of the Wilson loop is given by
\begin{equation}
\left\langle W \left( \mathcal{C} \right) \right\rangle \simeq e^{-S_{on-shell}}, \label{eq_WL2}
\end{equation}
where $S_{on-shell}$ is the on-shell string action on a world-sheet bounded by a loop $\mathcal{C}$ at the boundary of AdS space, which is proportional to the minimum area of the string world-sheet. Comparing Eq.(\ref{eq_WL1}) and Eq.(\ref{eq_WL2}), the free energy of the meson can be calculated as
\begin{equation}
V_{q\bar{q}}(r,T) = T S_{on-shell}(r,T). \label{eq_V}
\end{equation}

\subsection{Configurations of Open Strings}
\begin{figure}[t!]
\begin{center}
\includegraphics[
height=2.6in, width=3.6in]
{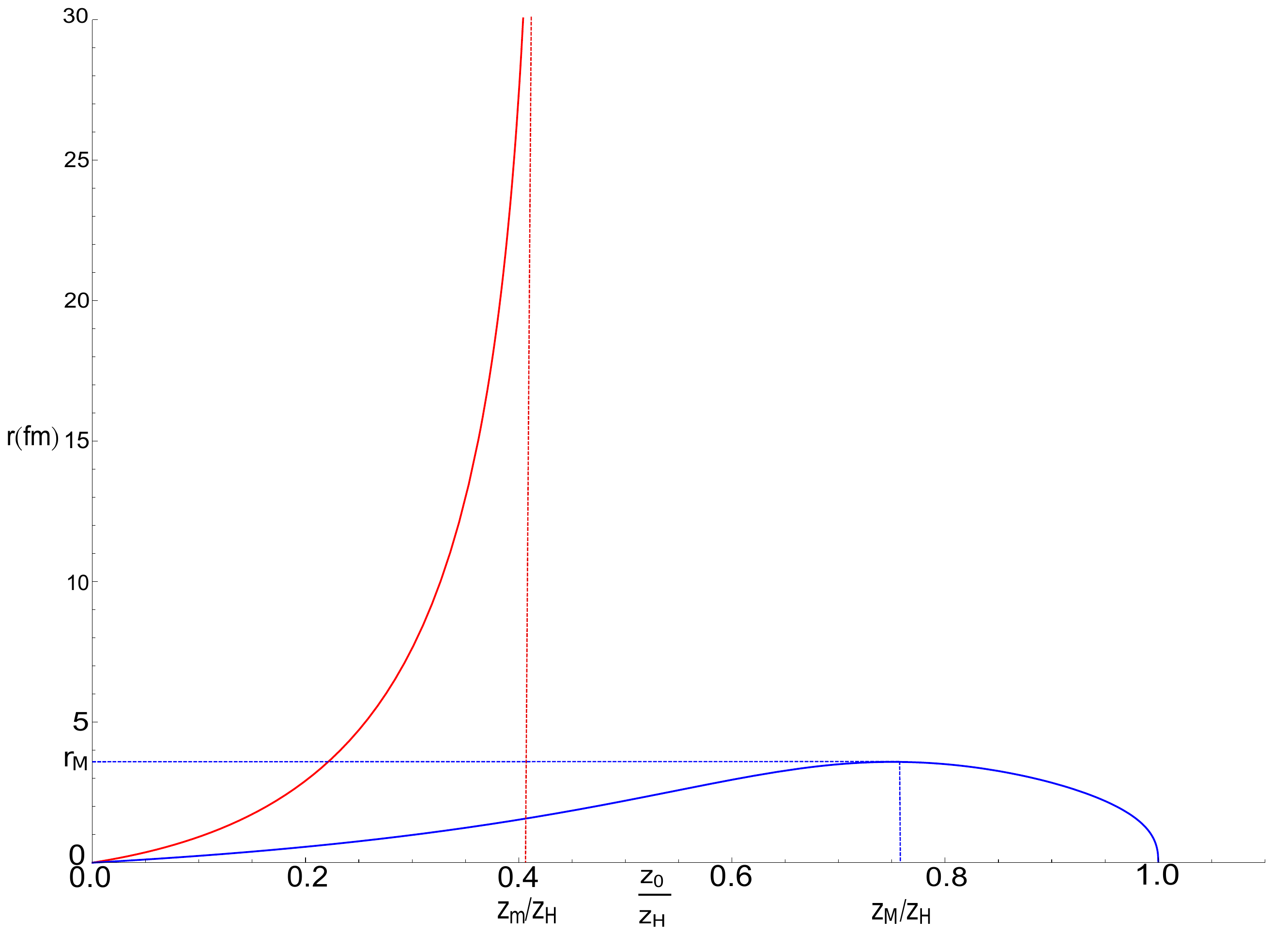}
\end{center}
\caption{The plot of the distance between quark and antiquark $r$ v.s. the maximum depth $z_0$ of the open string in U-shape. The red and blue lines are for $z_H=9.0$ and $z_H=3.5$ at $\mu=0.12$, which represents the small and large black hole respectively.} \label{fig_rz0}
\end{figure}

The string world-sheet action is defined in Eq.(\ref{SNG}) with the induced metric on the string world-sheet in Eq.(\ref{IM}). For the U-shape configuration, by choosing static gauge: $\xi^{0}=t,$ $\xi^{1}=x$, the induced metric in string frame become
\begin{equation}
ds^{2}=G_{ab}d\xi^{a}d\xi^{b}
=\frac{e^{2A_s(z)}}{z^2}g(z) dt^2
 + \frac{e^{2A_s(z)}}{z^2}\left(1+\frac{z'^2}{g(z)}\right) dx^2,
\end{equation}
where the prime denotes a derivative with respect to $x$. The Lagrangian and Hamiltonian can be calculated as
\begin{equation}
\mathcal{L} =\frac{e^{2A_s(z)}}{z^{2}}\sqrt{g(z)+z'^2},~
\mathcal{H} =-\frac{e^{2A_s(z)}}{z^{2}}\frac{g(z)}{\sqrt{g(z)+z'^2}} \label{eq_H}.
\end{equation}
Given the boundary conditions
\begin{equation}
z\left( x=\pm\frac{r}{2} \right)=0,~z(x=0)=z_{0},~z'(x=0)=0,
\end{equation}
we obtain the conserved energy from Hamiltonian in Eq.(\ref{eq_H})
\begin{equation}
\mathcal{H}(x=0)=-\frac{e^{2A_s(z_0)}}{z_{0}^{2}}\sqrt{g(z_0)}.
\end{equation}
Therefore, the U-shape configuration of an open string can be solve by
\begin{equation}
z'=\sqrt{ g \left( \frac{\sigma^2(z)}{\sigma^2(z_0)}-1 \right) },
\end{equation}
where $\sigma$ is the effective string tension \cite{0611304},
\begin{equation}
\sigma(z)=\frac{e^{2A_s(z)}\sqrt{g(z)}}{z^2} \label{string tension}
\end{equation}
and the warped factor in string frame becomes
\begin{equation}
    A_s(z)=A(z)+\sqrt{\frac{1}{6}}\phi(z). \label{eq_As}
\end{equation}
The distance $r$ between the pair of quark-antiquark can be calculated as,
\begin{equation}
r=\int_{-\frac{r}{2}}^{\frac{r}{2}} dx
 =2\int_{0}^{z_0} dz \frac{1}{z'}
 =2\int_{0}^{z_0} dz
	\left[ g(z) \left( \frac{\sigma^2(z)}{\sigma^2(z_0)}-1\right) \right]^{\frac{1}{2}},
\end{equation}
where $z_{0}$ is the maximum depth that the string can reach. The dependence of the distance $r$ on $z_{0}$ at two different cases are plotted in Fig.\ref{fig_rz0}. The red (upper) line corresponds to the case of small black hole horizon that the open string reaches a maximum depth $z_m$ when $r\rightarrow \infty$ and can not reach the horizon, while the blue (lower) line corresponds to the case of large black hole horizon that the open string might reach the horizon but with a limited separation $r\leq r_M$.

\subsection{Cornell Potential}
The potential between a pair of quark-antiquark $V_{q\bar{q}}$ in Eq.(\ref{eq_V}) for the open strings in U-shape configurations can be calculated as
\begin{equation}
V_{q\bar{q}}=T S_{on-shell}
=\int_{-\frac{r}{2}}^{\frac{r}{2}} dx \mathcal{L}
=2\int_0^{z_0} dz \frac{\sigma(z)}{\sqrt{g(z)}}
	\left[ 1-\frac{\sigma^2(z_0)}{\sigma^2(z)}\right]^{-\frac{1}{2}}. \label{eq_CornellV}
\end{equation}
It is well known that the potential $V_{q\bar{q}}$ can be expressed in the form of Cornell potential, which behaves as Coulomb type in short separation between quark and antiquark, but linear behavior in large separation with the coefficient $\sigma_s$, the string tension,
\begin{equation}
V_{q\bar{q}}=-\frac{\kappa}{r}+\sigma_s r +C. \label{eq_VC}
\end{equation}
As, $r \to 0$, i.e. $z_0 \to 0$, we expand the distance $r$ and the potential $V_{q\bar{q}}$ at $z_{0}=0$,
\begin{eqnarray}
r &=& 2\int_0^{z_0} dz
		\left[g(z) \left(
			\frac{\sigma^{2}(z)}{\sigma^{2}(z_0)}-1
		\right) \right] ^{-\frac{1}{2}}
   =r_{1}z_{0}+O(z_{0}^{2}),\\
V_{q\bar{q}} &=& 2\int_0^{z_0} dz \frac{\sigma(z)}{\sqrt{g(z)}}
		\left[ 1-\frac{\sigma^{2}(z_0)}{\sigma^{2}(z)}\right]^{-\frac{1}{2}}
   =\frac{V_{-1}}{z_0}+O(1),
\end{eqnarray}
where  \footnote{We require the property of Beta function $\frac{B(x/k,y)}{k}=\int_0^1 dt~t^{x-1} (1-t^k)^{y-1}.$}
\begin{eqnarray}
r_{1} &=& 2\int_{0}^{1}dv \left( \frac{1}{v^{4}} -1 \right)^{-\frac{1}{2}}
=\frac{1}{2}B\left( \frac{3}{4},\frac{1}{2} \right), \\
V_{-1} &=& 2\int_{0}^{1} \frac{dv}{v^{2}} \left( 1-v^{4} \right)^{-\frac{1}{2}}
=\frac{1}{2}B\left( -\frac{1}{4},\frac{1}{2} \right),
\end{eqnarray}
which gives the Coulomb potential,
\begin{equation}
V_{q\bar{q}}=\frac{r_1 V_{-1}}{r}+...,
\end{equation}
with the coefficient
\begin{equation}
\kappa = - r_1 V_{-1} \simeq 1.4355.\label{Coulomb coefficient}
\end{equation}
As $r \to \infty$, i.e. $z_{0} \to z_m$, we make a coordinate transformation $z=z_{0}-z_{0}w^{2}$. The distance $r$ and the potential $V_{q\bar{q}}$ become
\begin{equation}
r = 2\int_0^{1} f_{r}(w) dw, ~~V = 2\int_0^{1} f_{V}(w) dw, \label{rV}
\end{equation}
where
\begin{align}
f_{r}\left(  w\right)   &  =2z_{0}w\left[  g\left(  z_{0}-z_{0}w^{2}\right)
\left(  \dfrac{\sigma^{2}\left(  z_{0}-z_{0}w^{2}\right)  }{\sigma^{2}\left(
z_{0}\right)  }-1\right)  \right]  ^{-\frac{1}{2}},\\
f_{V}\left(  w\right)   &  =2z_{0}w\frac{\sigma\left(  z_{0}-z_{0}%
w^{2}\right)  }{\sqrt{g\left(  z_{0}-z_{0}w^{2}\right)  }}\left[
1-\dfrac{\sigma^{2}\left(  z_{0}\right)  }{\sigma^{2}\left(  z_{0}-z_{0}%
w^{2}\right)  }\right]  ^{-\frac{1}{2}}.
\end{align}
From Fig.\ref{fig_rz0}, the distance $r$ is divergent at $z_{0}=z_{m}$. By carefully analysis, we find that this divergence also happens for the quark potential because both the integrands $f_{r}(w)$ and $f_{V}(w)$ are divergent near the lower limit $w=0$, i.e. $z=z_{0}\rightarrow z_{m}$. To study the behaviours of distance $r$ and potential $V_{q\bar{q}}$ near $z_{0}=z_{m}$, we expand $f_r(w)$ and $f_V(w)$ at $w=0$,
\begin{eqnarray}
f_r(w) &=& 2 z_0 \left[ -2 z_0 g(z_0)
	\frac{\sigma'(z_0)}{\sigma(z_0)}\right]^{-\frac{1}{2}}+O(w), \label{eq_fr} \\
f_V(w) &=& 2 z_0 \sigma(z_0)
	\left[-2 z_0 g(z_0)\frac{\sigma'(z_0)}{\sigma(z_0)}\right]^{-\frac{1}{2}}+O(w). \label{eq_fV}
\end{eqnarray}
\begin{figure}[t!]
\begin{center}
\includegraphics[
height=2.6in, width=3.6in]
{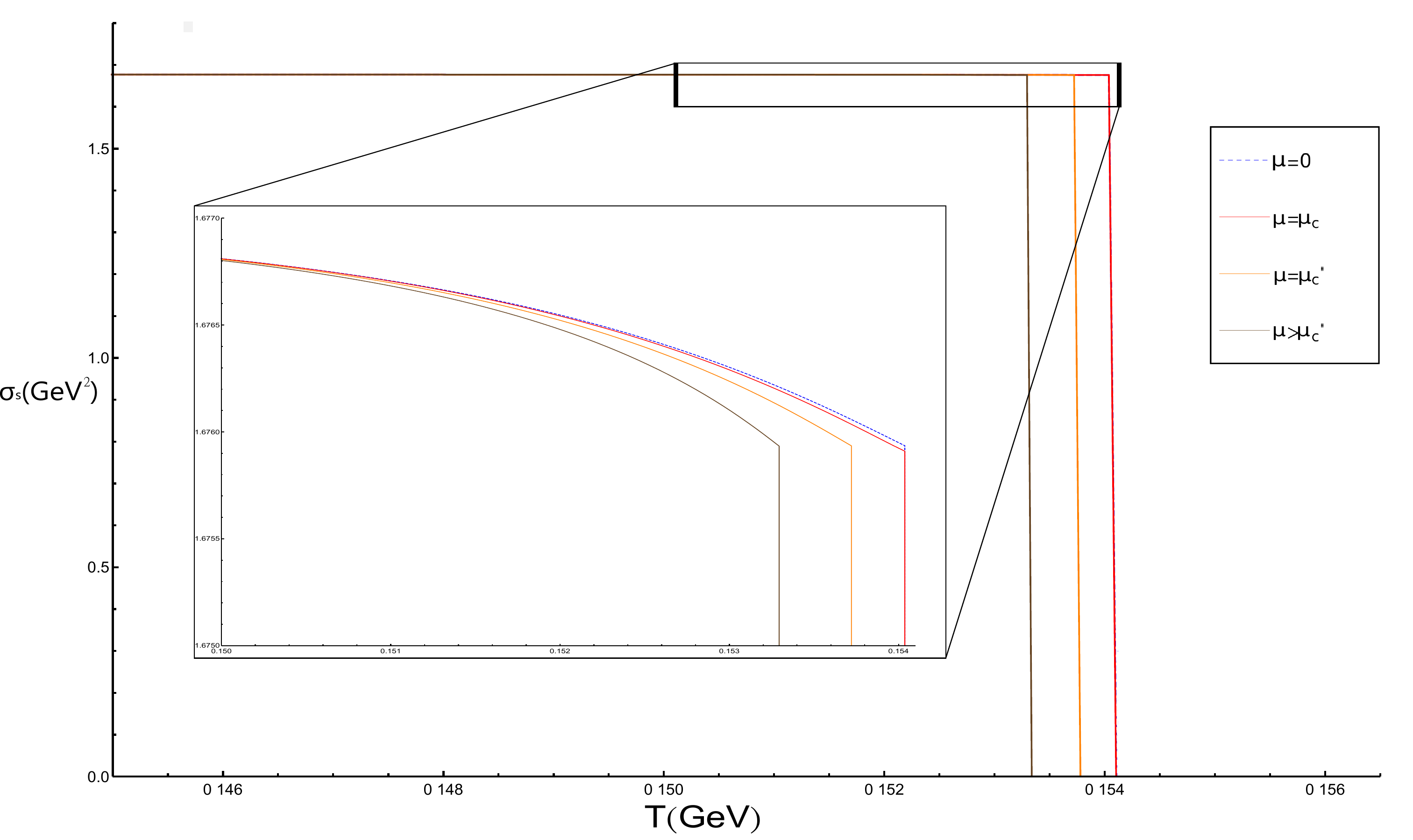}
\end{center}
\caption{The string tension $\sigma_s$ v.s temperature in different chemical potentials with $\mu_c=0.04779$ and $\mu_c'=0.1043$. The region near the phase transition is enlarged, where the tension suddenly drop to zero implying the phase transition between confinement and deconfinement phases.} \label{fig_sigmasT}
\end{figure}
The integrals in Eq.(\ref{rV}) can be approximated by only considering the leading terms of $f_r(w)$ and $f_r(w)$ near $z_0=z_m$ in Eqs.(\ref{eq_fr}-\ref{eq_fV}). This leads to
\begin{eqnarray}
r(z_0) &\simeq & 4z_0
	\left[ -2z_0 g(z_0) \frac{\sigma'(z_0)}{\sigma(z_0)}\right]
		^{-\frac{1}{2}},\\
V(z_0) &\simeq & 4z_0 \sigma(z_0)
	\left[ -2z_0 g(z_0) \frac{\sigma'(z_0)}{\sigma(z_0)}\right]
		^{-\frac{1}{2}}
	=\sigma(z_0) r(z_0).
\end{eqnarray}
From the above expression, we obtain the expected linear potential $V=\sigma_{s}r$ at long distance with the string tension,
\begin{equation}
\sigma_{s}=\left.  \dfrac{dV}{dr}\right\vert_{z_{0}=z_{m}}=\left.
\dfrac{dV/dz_{0}}{dr/dz_{0}}\right\vert _{z_{0}=z_{m}}=\left.  \dfrac
{\sigma^{\prime}\left( z_{0}\right) r\left( z_{0}\right)  +\sigma\left(z_{0}\right) r^{\prime}\left( z_{0}\right) }{r^{\prime}\left(z_{0}\right) }\right\vert _{z_{0}=z_{m}}=\sigma\left( z_{m}\right) .
\end{equation}
The temperature dependence of the string tension for various chemical potentials is plotted in Fig.\ref{fig_sigmasT}. We see that the string tension decreases when the temperature increases. At the confinement-deconfinement transformation temperature $T_{\mu}$, the system transforms to the deconfinement phase and the string tension suddenly drops to zero as we expected \cite{1006.0055}. The behaviour is consistent with the result of lattice QCD simulation \cite{1006.0055}. In Fig.\ref{fig_sigmasT}, $\mu_c=0.04779$ is the critical chemical potential for the black holes phase transition in background and $\mu_c'=0.1043$ is the critical chemical potential for confinement-deconfinement phase transition in our holographic QCD model. We will discuss these two phase transitions in detail in the next section.

We therefore showed that the behaviours of the quark potential at short distance and long distance agrees with the form of the Cornell potential \cite{Cornell},
\begin{equation}
V\left( r\right) = -\dfrac{\kappa}{r}+\sigma_{s} r+C, \label{Cornell}
\end{equation}
which has been measured in great detail in lattice simulations \cite{2001}.
\begin{figure}[t!]
\begin{center}
\includegraphics[
height=2.4in, width=3.6in]
{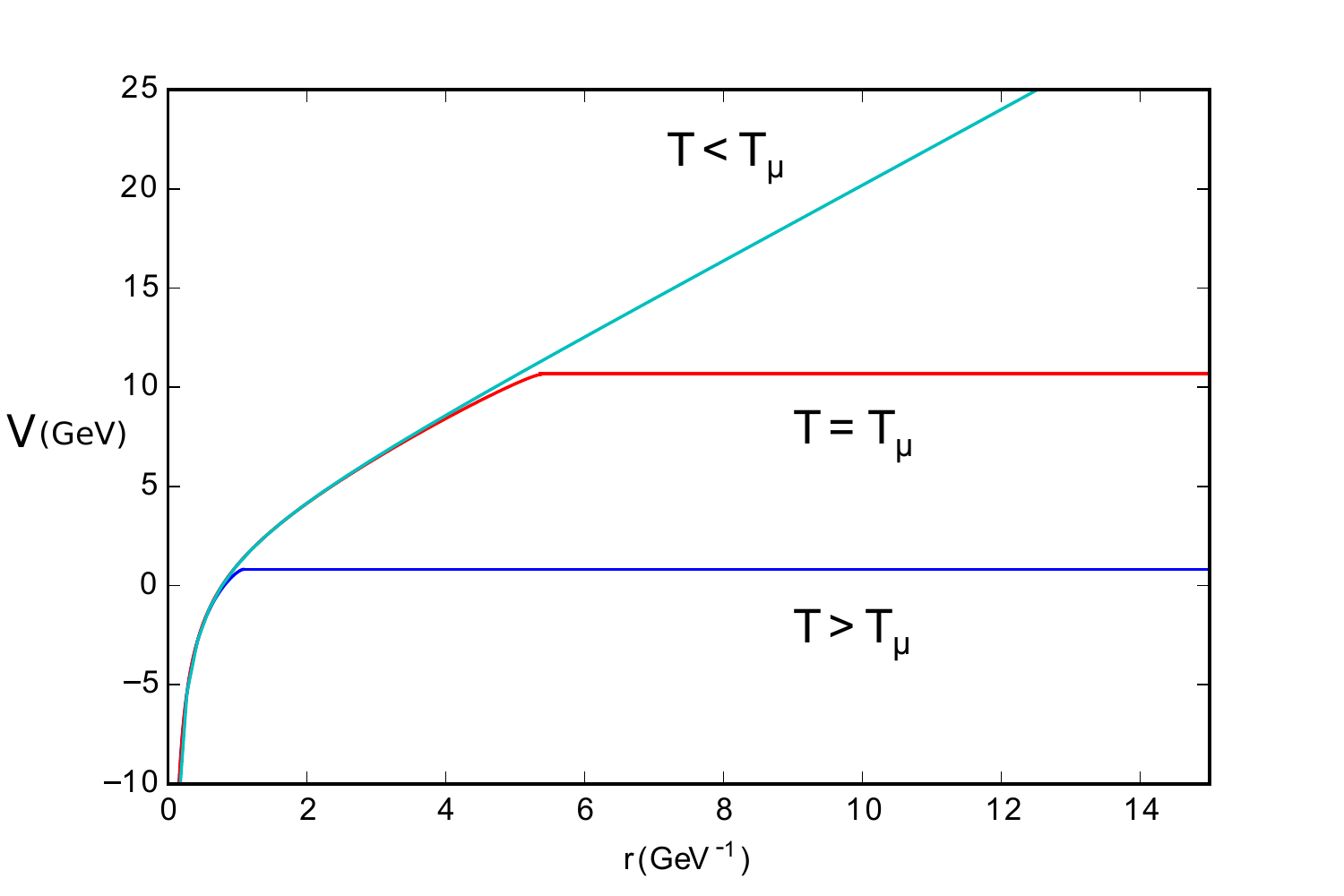}
\end{center}
\caption{The regularized heavy quark potential $V_{q\bar{q}}^{[R]}$ v.s. $r$ at $\mu=0.12$ GeV for $z_H=\{1,3,10\}$ from lower to upper lines} \label{fig_Vr}
\end{figure}
In order to obtain the $r$ dependence of $V_{q\bar{q}}$, we will evaluate the integral in Eq.(\ref{eq_VC}), which is divergence due to integrand is not well defined at $z=0$. We regularize $V_{q\bar{q}}$ by subtracting the divergent terms as
\begin{equation}
V_{q\bar{q}}^{[R]}=C(z_{0}) + 2\int_{0}^{z_0}dz\left[ \frac{\sigma(z)}{\sqrt{g(z)}}\left[ 1-\frac{\sigma^{2}(z_0)}{\sigma^{2}(z)}\right]^{-\frac{1}{2}}-\frac{1}{z^{2}}\left[ 1+2A'(0) z\right] \right] , \label{eq_VR}
\end{equation}
where
\begin{equation}
C(z_0) = -\frac{2}{z_{0}}+4A_s'(0)\ln z_{0}.
\end{equation}
The regularized potentials are plotted in Fig.\ref{fig_Vr}. For $T<T_\mu$, $V_{q\bar{q}}^{[R]}$ has the form of Cornell potential with linear behavior for large $r$. For $T>T_\mu$, the open string breaks at certain distant and $V_{q\bar{q}}^{[R]}$ become constant for the larger distance.

\setcounter{equation}{0}
\renewcommand{\theequation}{\arabic{section}.\arabic{equation}}%

\section{Phase Diagram}
\begin{figure}[t!]
\begin{center}
\includegraphics[
height=2.0in, width=3.6in]
{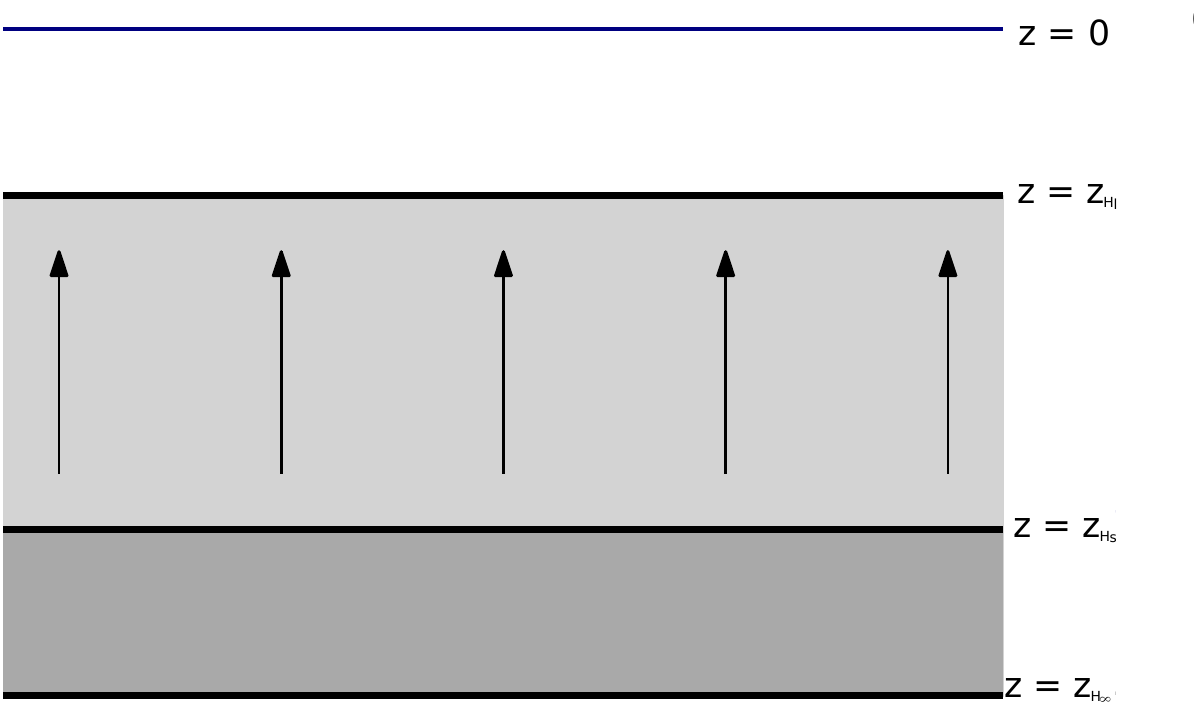}
\end{center}
\caption{The schematic diagram of the phase transition from small black hole $z_{H_{s}}$ into a large black hole $z_{H_{l}}$.} \label{fig_zhwall}
\end{figure}

In the previous sections, we have constructed a holographic QCD model by studying the Einstein-Maxwell-scalar system. We obtained a family of black hole backgrounds and studied the phase transition between the black holes by computing their free energies. We also added probe open strings in the black hole backgrounds and studied the different string configurations at various temperatures, which corresponding to the confinement and deconfinement phases in the dual holographic QCD model. In this section, we are ready to discuss the phase diagram of QCD by combining the phase structure of black hole background and the configurations of the probe open strings in the black hole background.
We have obtained the phase diagram for phase transitions of black hole background, as shown in Fig.\ref{fig_Tmu_BH}, by comparing the free energies of black holes. We summarize our results in a schematic diagram Fig.\ref{fig_zhwall}. As black hole temperature grows up, black hole horizon grows up as well, i.e. $z_{H}$ decreases. At the phase transition temperature, the small black hole with horizon $z_{H_{s}}$ jumps to the large one with horizon $z_{H_{l}}$.
\subsection{Probe strings and Dynamical Wall}
To see the confinement-deconfinement phase transition, we added probe open strings into the black hole background. As shown in Fig.\ref{fig_rz0}, for a small black hole $z_{H_{s}}$, the separation of the bounded quark-antiquark pair $r$ can be as long as possible, but the depth of the string is limited by a maximum value $z_m$, which we call dynamical wall. In this case, the open strings are always connected in the U-shape to form a bounded states and the system is in the confinement phase. The dynamical wall is the crucial concept to understand the confinement-deconfinement phase transition in holographic QCD models. On the other side, for a large black hole $z_{H_{l}}$, as shown in Fig.\ref{fig_rz0}, the separation of the quark-antiquark pair is bounded by a maximum distance $r_M$ at the depth $z_M$. If the distance between the quark and antiquark is longer than $r_M$, an U-shaped string will break into two I-shaped open strings attached between the boundary and the horizon. In this case, free quark or antiquark might exist indicating the system is in the deconfinement phase. The open string breaking process corresponds to the melting of the bounded state \cite{1605.07181}.

\begin{figure}[t!]
\begin{center}
\includegraphics[
height=1.8in, width=3in]
{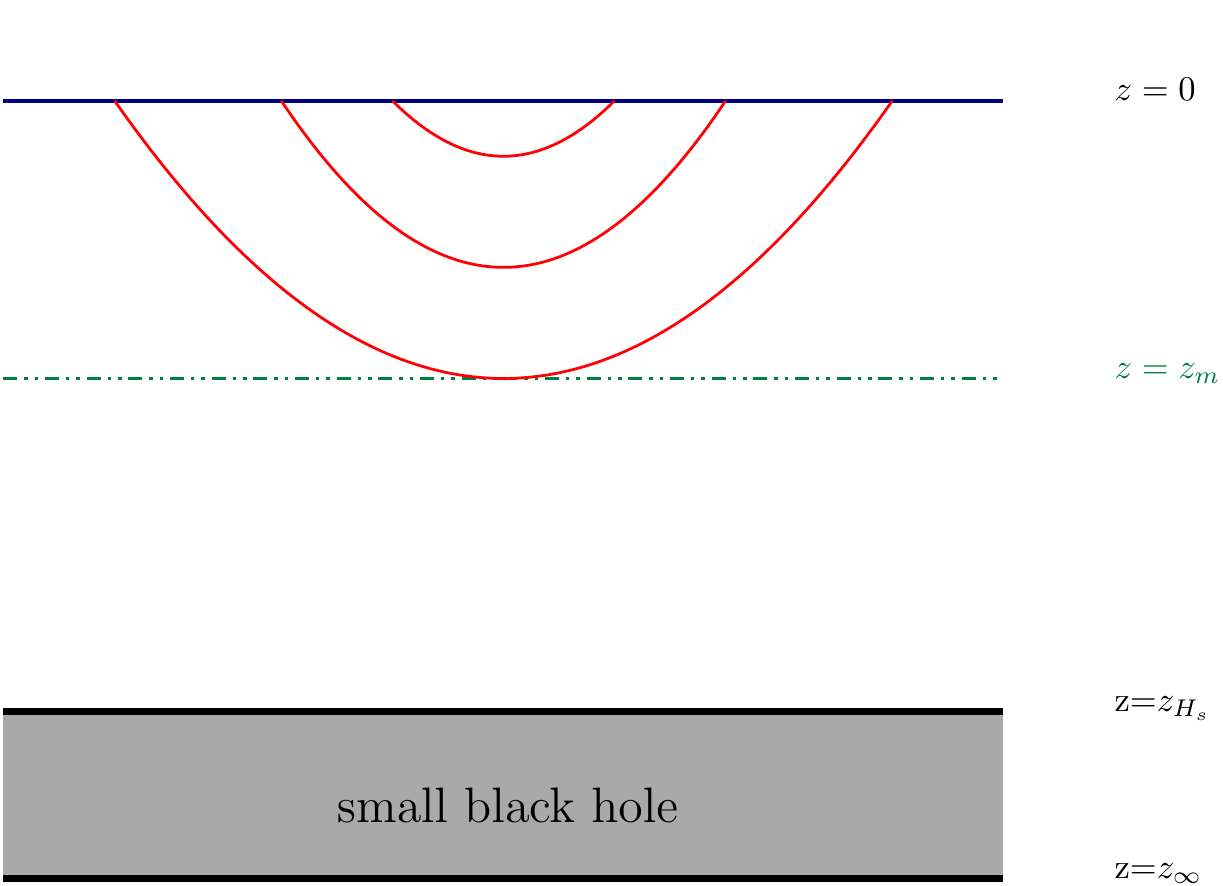}
\hspace*{0.5cm}
\includegraphics[
height=1.8in, width=3in]
{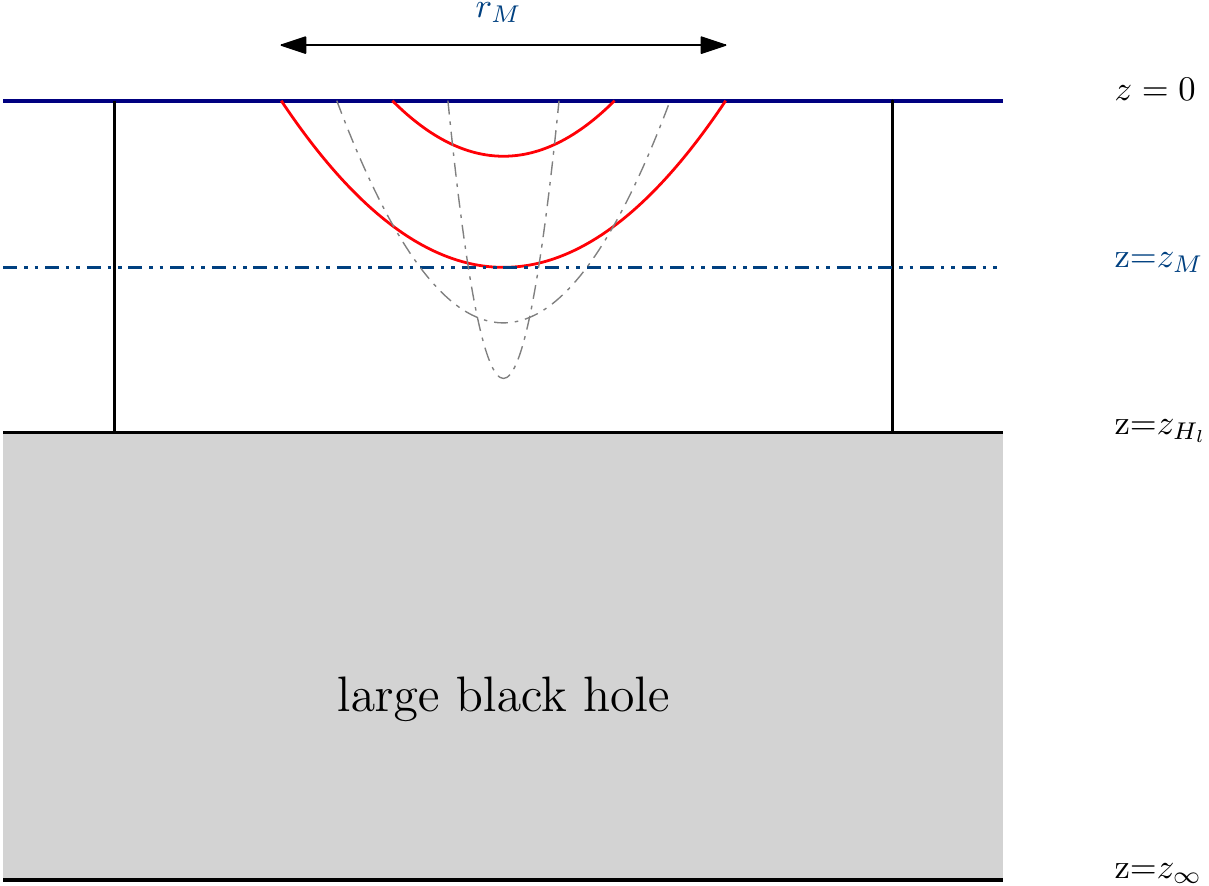}
\vskip -0.05cm \hskip -1.2 cm (a) \hskip 6.8 cm (b)
\end{center}
\caption{Small/large black hole in horizon description, $z_{H_{s/l}}$.} \label{fig_sLBH}
\end{figure}

We summarize our discussion in a schematic diagram Fig.\ref{fig_sLBH}. For a small black hole, as shown in Fig.\ref{fig_sLBH}(a), there exists a dynamical wall so that open strings are always in the U-shape corresponding to the confinement phase. While for a large black hole, as shown in Fig.\ref{fig_sLBH}(b), the open strings could be either U-shaped or I-shaped depending on the separating distant between the quark and antiquark corresponding to the deconfinement phase. The configurations of open strings are collected in table 2. Since the black hole temperature are closely associated to the black hole horizon, we expect that the system will undergo a phase transition from confinement to deconfinement when temperature increases.

Since the role of dynamical wall is crucial to affirm the confinement phase in holographic QCD models, let us examine it carefully. To determine the position of the dynamical wall $z_m$, we use the fact that $r(z_m)\rightarrow\infty$, which leads to the equation $\sigma'(z_m)=0$. With the definition of string tension in Eq.(\ref{string tension}), we \begin{equation}
\frac{g'(z_m)}{4g(z_m)}+A'_s(z_m)+\frac{1}{z_m}=0.
\end{equation}
In the confinement phase, the value of horizon $z_H$ is large, so that $g(z)$ is almost a constant and $g'(z)\sim 0$, we thus have
\begin{equation}
A_s'(z_m)+\frac{1}{z_m}=0.
\end{equation}
We would like to remark that the position of the dynamical wall $z_m$ is almost an universal value depending neither on the chemical potential nor on the temperature\footnote{Here we mean that the system in the confinement phase. In deconfinement phase, there is no dynamical wall.}. In our particular model with the choice of $A(z)$ in Eq.(\ref{eq-Ae}) and also string frame $A_s$ in Eq.(\ref{eq_As}), the position of the dynamical wall $z_m$ can be obtained as
\begin{equation}
z_m=\sqrt{\frac{a-1-\sqrt{a(a-4)}}{b(2a+1)}}\simeq 4.22.\label{zm}
\end{equation}
\begin{table}[t!]
\begin{center}
\begin{tabular}
[c]{|c|c|c|c|}\hline
Black hole size         & U-shape      & I-shape & Phase in QCD\\\hline
Small $BH (z_{H_{s}})$ & $0<r<\infty$ & none    & Confinement\\\hline
Large $BH (z_{H_{l}})$ & $r<r_M$      & $r>r_M$ & Deconfenment\\\hline
\end{tabular}
\end{center}
\caption{Black holes sizes and open strings configurations with respect to the $q\bar{q}$ separation $r$.}
\label{tab_cfgn_r}
\end{table}

\begin{figure}[t!]
\begin{center}
\includegraphics[
height=2.2in, width=3.2in]
{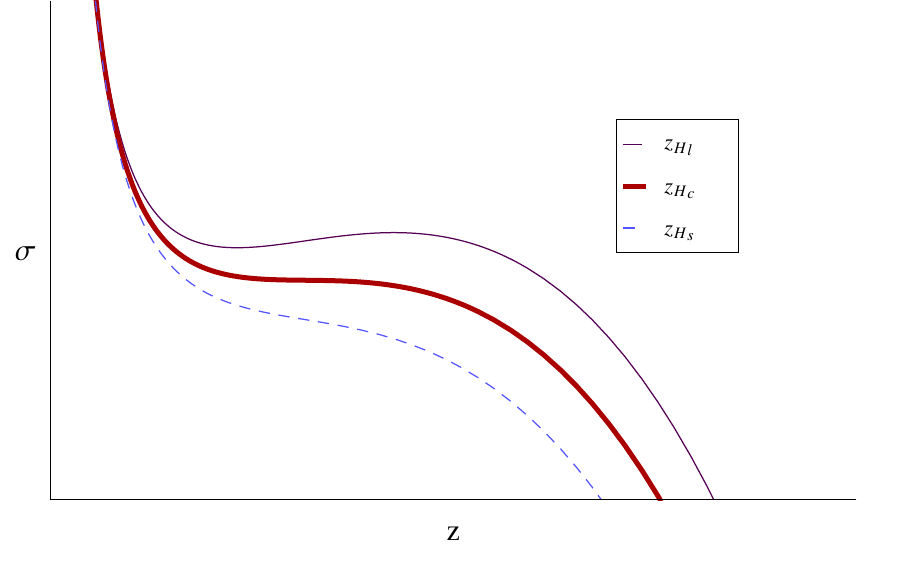}
\includegraphics[
height=2.2in, width=3.2in]
{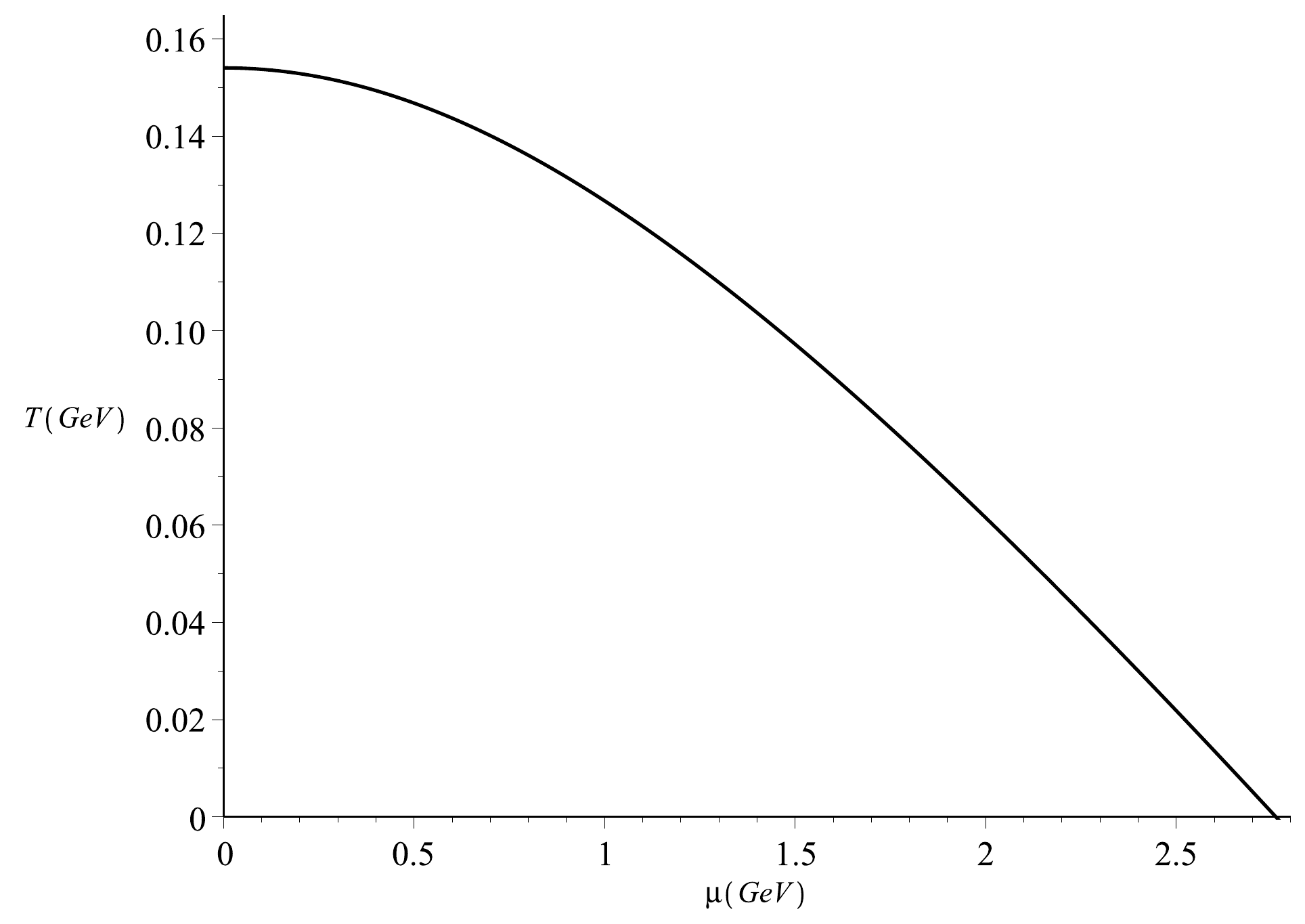}
\vskip -0.05cm \hskip -1.2 cm (a) \hskip 6.8 cm (b)
\end{center}
\caption{(a) The effective string tension $\sigma$ for different temperatures. (b) The transformation temperature $T_\mu$, associating to $z_{H_{\mu}}$, at which the dynamical wall appears/disappears for each chemical potential. Where the $T$-intercept, $T_0=0.1541$ GeV.} \label{fig_sigmaz} \label{fig_Tmu_string}
\end{figure}

For each chemical potential $\mu$, we define the transformation temperature $T_{\mu}$ corresponding to the critical black hole horizon $z_{H_\mu}$, at which the dynamical wall appears/disappears as shown in Fig.\ref{fig_Tmu_string}(a). The transformation temperature is associated to the transformation between the confinement and deconfinement phases in the holographic QCD. The transformation temperature $T_\mu$ at each chemical potential is plotted in Fig.\ref{fig_Tmu_string}(b). We should emphasize that the transformation between the confinement and deconfinement phases here is not a phase transition. The confinement phase smoothly transforms to the deconfinement phase as the temperature increases gradually.
\begin{figure}[t!]
\begin{center}
\includegraphics[
height=2.2in, width=3.2in]
{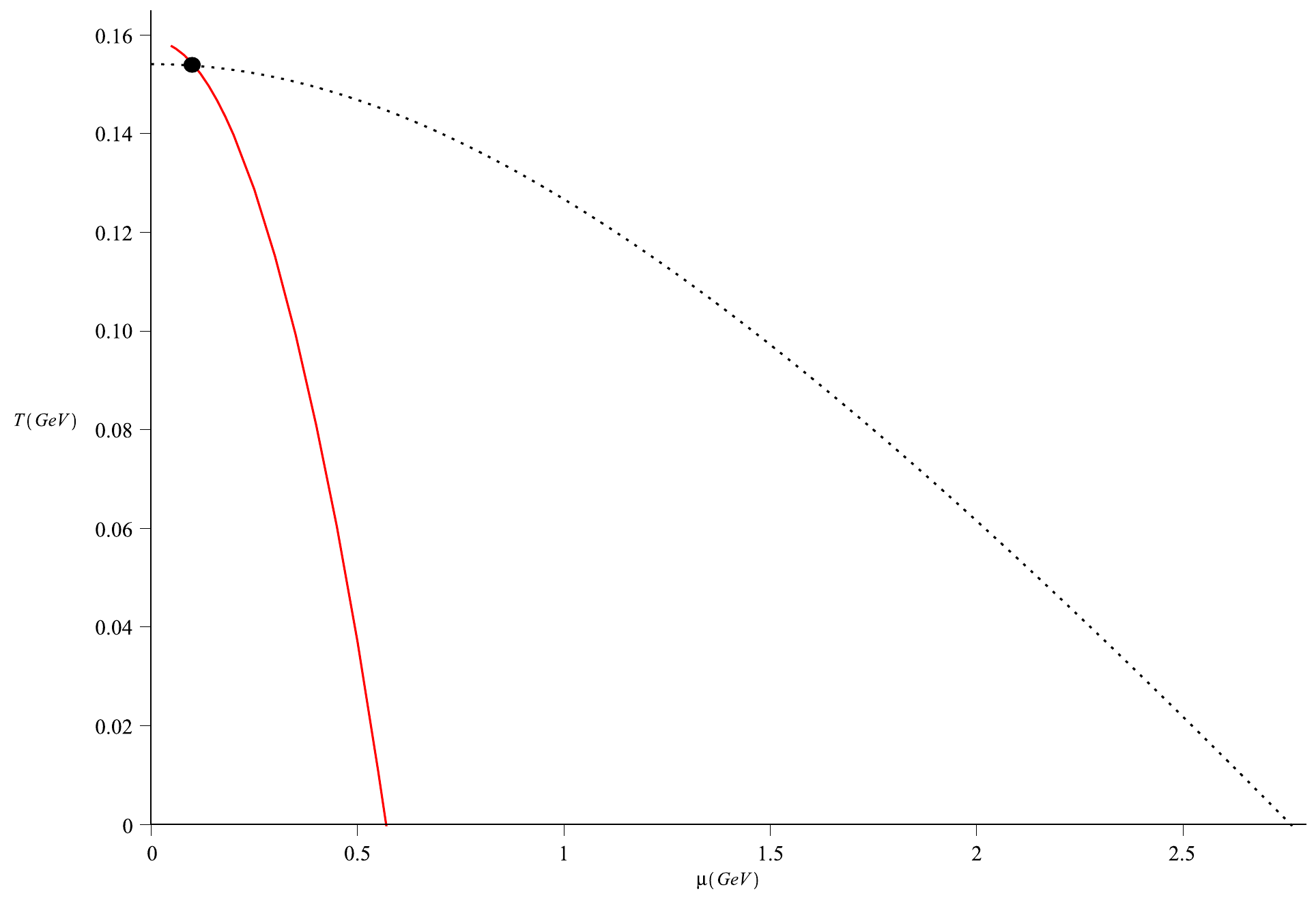}
\includegraphics[
height=2.2in, width=3.2in]
{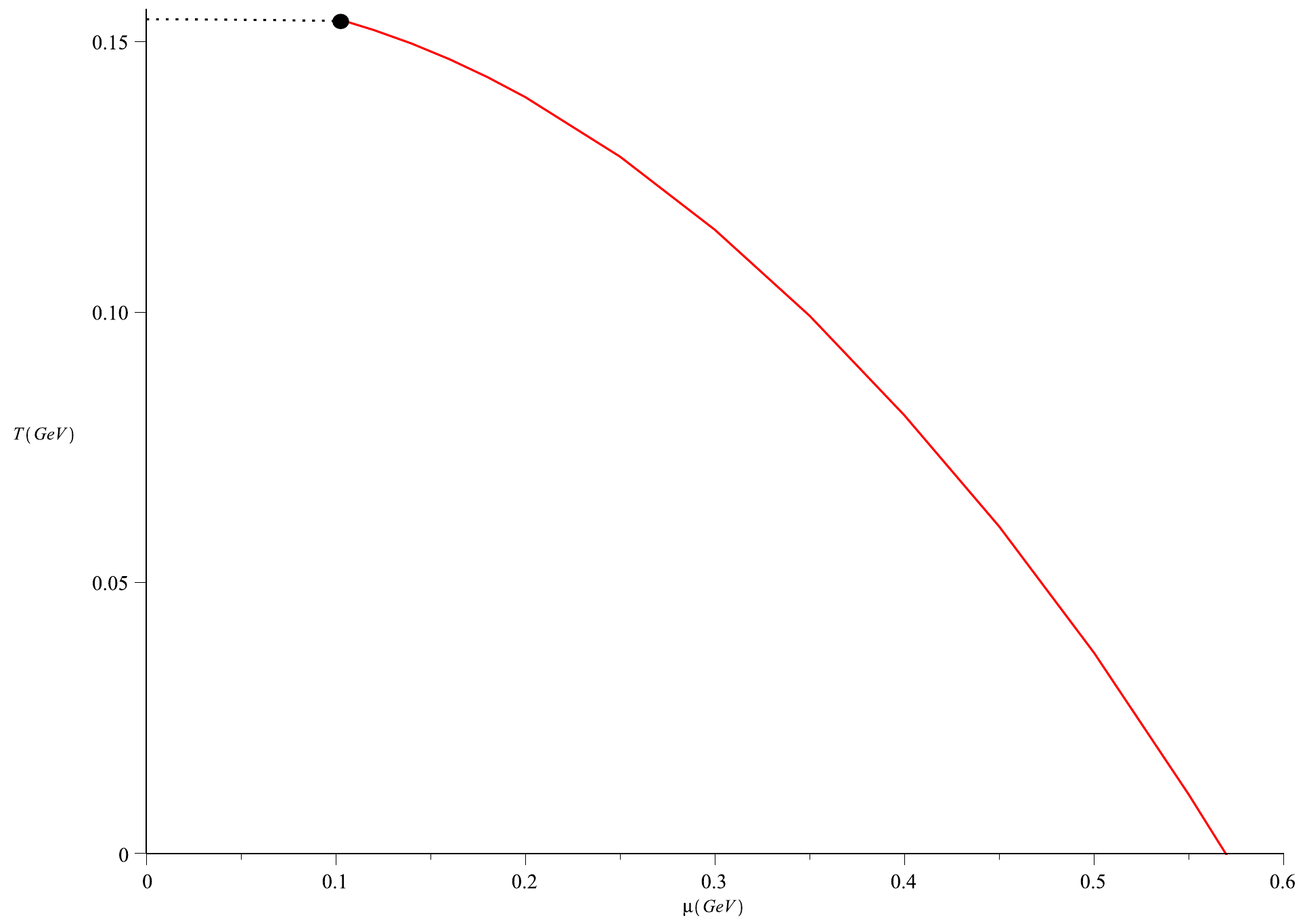}
\vskip -0.05cm \hskip -1.2 cm (a) \hskip 6.8 cm (b)
\end{center}
\caption{The phase diagram of a probing string in a black hole system and the intersection is at $(\mu_c'=0.1043,T_c'=0.1538)$GeV.} \label{fig_Tmuboth}
\end{figure}

\subsection{Confinement-deconfinement Phase Diagram}
To obtain the completed phase diagram in our holographic QCD model, we combine the phase transition between large and small black holes as well as the different configurations of the probe open strings in the background. In Fig.\ref{fig_Tmuboth}(a), the black (dotted) line represents the confinement-deconfinement transformation, while the red (solid) line is the phase transition between black holes. Once the phase transition from a small black hole at lower temperature to a large black hole at higher temperature takes place, the black hole horizon jumps to a large value which is beyond the critical black hole horizon $zH_\mu$ and the system performs the confinement-deconfinement phase transition. The intersection of two lines is identified as the critical point at $(\mu_c'=0.1043,T_c'=0.1538)$, which is consistent with the recent result by lattice QCD in \cite{1701.04325}.

The final phase diagram of the confinement-deconfinement phase transition is plotted in \ref{fig_Tmuboth}(b). For a large chemical potential $\mu>\mu_c'$, there is a first order confinement-deconfinement phase transition at the temperature $T_\mu$ shown as solid red line. At the critical point $\mu=\mu_c'$ and $T=T_c'$ shown as a black dot, the first order phase transition weakens to a second order phase transition. For small chemical potential $\mu<\mu_c'$, the confinement-deconfinement phase transition reduces to the smooth crossover shown as the dotted black line.

\section{Conclusion}
In this paper, we constructed a bottom-up holographic QCD model by studying gravity coupled to a $U(1)$ gauge field and a neutral scalar in five dimensional space-time, i.e. 5-dimensional Einstein-Maxwell-scalar system. By solved the equations of motion analytically, we obtained a family of black hole solutions which depend on two arbitrary functions $f(z)$ and $A(z)$. Different choices of the functions $f(z)$ and $A(z)$ corresponds to different black hole backgrounds. To include meson fields in QCD, probe gauge fields were added on the 5-dimensional backgrounds. The function $f(z)$ can be fixed by requiring the linear Regge spectrum for mesons. By a suitable choice of the function $A(z)$ as in Eq.(\ref{eq-Ae}), we fixed our holographic QCD model.

We obtained the phase structure of the black hole background by studying its thermodynamics quantities. To realize the confinement-deconfinement phase transition in QCD, we added probe open strings in the black hole background and studied their stable configurations of shape. Different configurations of U-shape and I-shape were identified to confinement and deconfinement phases, respectively. By combining the phase structure of the black hole background and the U-shape to I-shape transformation of the open strings, we obtain the phase diagram of confinement-deconfinement phase transition in our holographic QCD model. In our model, the critical point, where the first order phase transition becomes crossover, is predicted at $(0.1043~GeV,0.1538~GeV)$ which is consistent with the recent result of lattice QCD result in \cite{1701.04325}.

\begin{figure}[t!]
\begin{center}
\includegraphics[
height=2.5in, width=3.6in]
{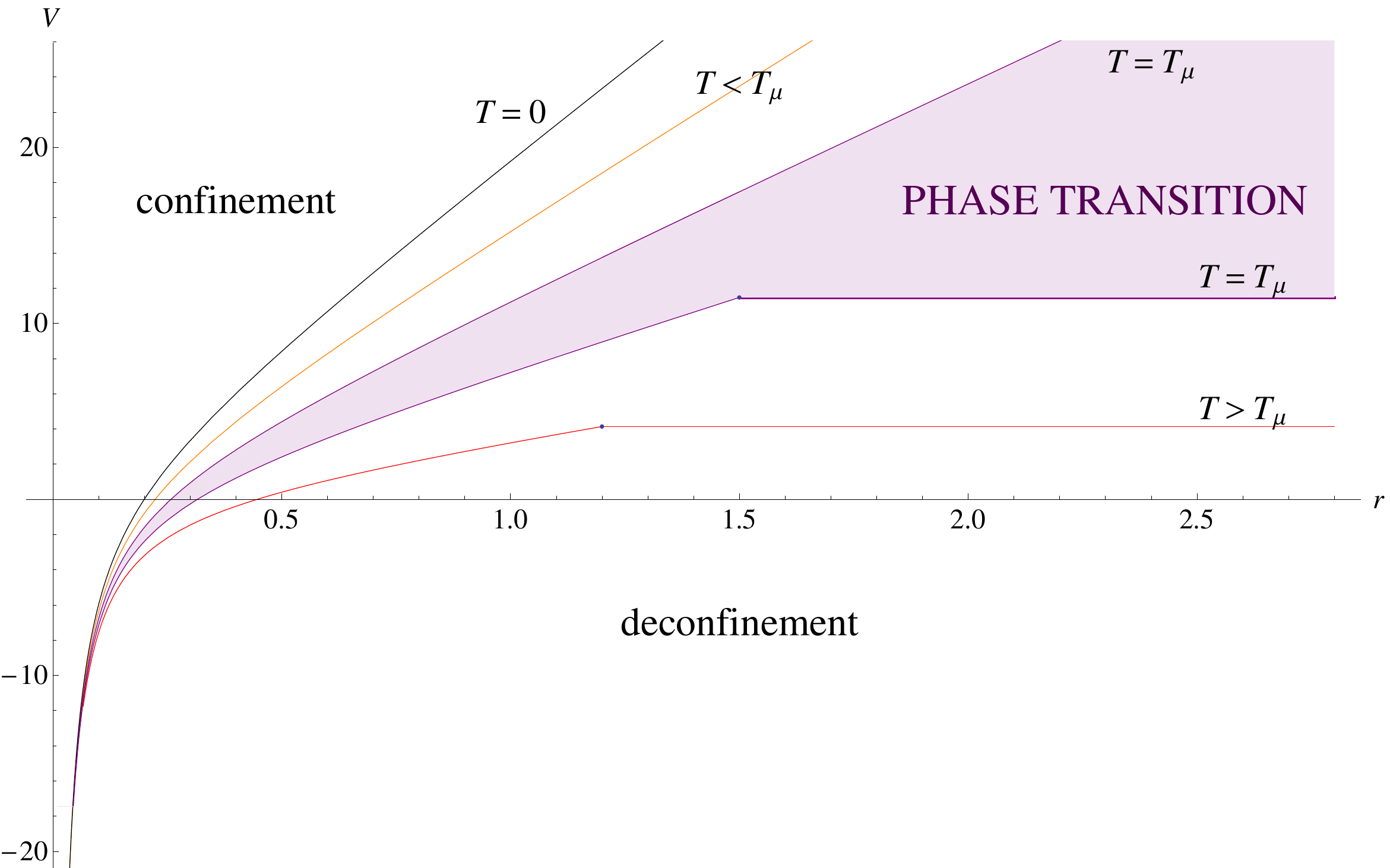}
\end{center}
\caption{The heavy quark potential at various temperatures. } \label{fig_cornell0}
\end{figure}

We studied Wilson loop in QCD by calculating the world-sheet area of an open string based on the holographic correspondence. The heavy quark potential can be obtained from the Wilson loop. We obtained the Cornell potential which has been well studied by lattice QCD. The sketch diagram of the heavy quark potentials at different temperatures is plotted in Fig.\ref{fig_cornell0}. At low temperature $T<T_\mu$, the potential is linear at large $r$ corresponding to the confinement phase; While at high temperature $T>T_\mu$, the potential becomes constant at large $r$ corresponding to the deconfinement phase. There is a phase transition at $T=T_\mu$ as it is showed in the sketch diagram.

The main properties in our holographic QCD model are summarized in the following:
\begin{itemize}
    \item The coupled Einstein-Maxwell-scalar system was solved analytically to obtain a family of black hole backgrounds in Eqs.(\ref{phip-A}-\ref{V-A}).
    \item The meson spectrum in our model satisfies the linear Regge behavior as in Eq.(\ref{mass}).
    \item For finite chemical potentials $\mu>\mu_c$, the background emerges a phase transition between a small black hole and a large black hole as shown in Fig.\ref{fig_Tmu_BH}(b).
    \item The dynamical wall appears/disappears for small/large black holes, which implies the confinement-deconfinement phase transition. In addition, in confinement phase, the position of the dynamical wall is nearly a constant as in Eq.(\ref{zm}) independent of the chemical potential and the temperature.
    \item We obtained the Cornell form of quark potential in Eq.(\ref{Cornell}) by calculating the Wilson loop.
\end{itemize}

\subsection*{Acknowledgements}

We would like to thank Rong-Gen Cai, Song He, Mei Huang, Danning Li, Xiaofeng Luo, Xiaoning Wu for useful
discussions. This work is supported by the Ministry of Science and Technology (MOST 105-2112-M-009-010) and National Center for Theoretical Science, Taiwan.

\end{document}